\documentclass[11pt,nofootinbib,preprint]{revtex4}
\usepackage{mathrsfs}
\usepackage{graphicx}
\usepackage{amsmath}
\usepackage{amsfonts}
\usepackage{amssymb}
\usepackage{color}

\usepackage{epsfig}
\usepackage{CJK}
\usepackage{graphicx}
\usepackage{epsfig}
\usepackage{eepic}
\usepackage{bbm}
\usepackage{dcolumn}
\usepackage{bm}
\usepackage{ulem}
\usepackage{slashbox}
\usepackage{multirow}
\usepackage{slashed}

\newcommand{\omits}[1]{}

\def\bc{\begin{center}}

\def\ec{\end{center}}
\def\be{\begin{eqnarray}}
\def\ee{\end{eqnarray}}

\definecolor{dyellow}{rgb}{1.,0.8,.0}
\definecolor{myblue}{rgb}{.1,.1,.7}
\definecolor{dcyan}{rgb}{.0,.6,.6}
\definecolor{cyan}{rgb}{0.4,1.0,1.0}
\definecolor{dmagenta}{rgb}{0.6,0.0,0.6}
\definecolor{brown}{rgb}{0.6,0.2,0.}
\definecolor{darkblue}{rgb}{.0,.0,0.5}
\definecolor{darkred}{rgb}{0.75,0.0,0.0}
\definecolor{orange}{rgb}{1.,.6,.0}
\definecolor{dorange}{rgb}{0.8,.4,.0}
\definecolor{green}{rgb}{0.0,1.0,0.0}
\definecolor{darkgreen}{rgb}{0.0,0.6,0.0}
\definecolor{purple}{rgb}{.4,.0,.4}
\definecolor{lightgrey}{rgb}{0.7, 0.7, 0.7}
\definecolor{grey}{rgb}{0.4, 0.4, 0.4}

\newcommand{\nc}{\newcommand}
\nc{\rnc}{\renewcommand} \nc{\ket}[1]{\left | \, #1 \right \rangle}
\nc{\bra}[1]{\left \langle #1 \, \right |}
\nc{\ua}{\uparrow} \nc{\da}{\downarrow}

\nc{\braket}[2]{\langle\, #1\,|\,#2\,\rangle}
\nc{\half}{\frac{1}{2}}

\nc{\prj}{\mathcal{P}} \nc{\hilb}{\mathcal{H}}
\nc{\pth}{\mathcal{C}} \nc{\inprod}[2]{\braket{#1}{#2}}
\nc{\upket}{\ket{\uparrow}} \nc{\downket}{\ket{\downarrow}}
\nc{\upbra}{\bra{\uparrow}} \nc{\downbra}{\bra{\downarrow}}

\begin{document}


\title{The PEE aspects of entanglement islands from bit threads}

\author{Yi-Yu Lin$^1$} \email{linyy27@mail2.sysu.edu.cn}
\author{Jia-Rui Sun$^{1}$} \email{sunjiarui@mail.sysu.edu.cn}
\author{Yuan Sun$^{2}$} \email{sunyuan@jlu.edu.cn}
\author{Jie-Chen Jin$^1$} \email{jinjch5@mail2.sysu.edu.cn}

\affiliation{${}^1$School of Physics and Astronomy, Sun Yat-Sen University, Guangzhou 510275, China}
\affiliation{${}^2$Center for Theoretical Physics and College of Physics, Jilin University,
Changchun 130012, China}



\begin{abstract}
We study the partial entanglement entropy (PEE) aspects of the holographic BCFT setup with an entanglement island, inspired by the holographic triality of the AdS/BCFT setup developed in the recent study on the black hole information problem, and the ``PEE=CFF (component flow flux)'' prescription, which is proposed recently to investigate the holographic PEE in the framework of bit thread formulation. Our study provides a bit thread description of the AdS/BCFT setup, which characterizes the specific entanglement details between the different parts of the system with an entanglement island, and may provide further insight into the black hole information problem. Furthermore, we show that in the context of island, one should distinguish between the fine-grained PEE and the semi-classical PEE. Interestingly, similar to the island rule of the fine-grained entropy in the semi-classical picture, we also propose the island rules of the fine-grained PEE.

\end{abstract}


\maketitle
\tableofcontents

\section{Introduction}

Recently, the AdS/BCFT (boundary conformal field theory) duality~\cite{Takayanagi:2011zk,Fujita:2011fp} has attracted lots of attention in the context of the black hole information problem~\cite{Page:1993wv,Page:2013dx} and quantum gravity, by combining with the so-called double holographic model~\cite{Almheiri:2019hni}. This stems from the interesting property of the holographic triality of this setup (see figure~\ref{f0}). According to this triality, the AdS/BCFT setup can be related to the double holographic model considered in recent studies on the black hole information paradox, in which an AdS black hole is coupled to an auxiliary, non-gravitational holographic CFT, often called a bath, which captures the Hawking radiation. It has been shown that although the complete theory of quantum gravity is a mystery still, one can describe the von Neumann entropy (or fine-grained entropy) of a subregion $R$ in the bath in the framework of semi-classical gravity picture. This is called the island prescription for the von Neumann entropy of Hawking radiation $R$~\cite{Almheiri:2019hni,Penington:2019npb,Almheiri:2019psf,Almheiri:2019qdq,Penington:2019kki}~\footnote{For further discussion on the lsland formula and black hole information problem see e.g.~\cite{Rozali:2019day,Sully:2020pza,Almheiri:2019yqk,Akers:2019nfi,Chen:2019uhq,Liu:2020gnp,Balasubramanian:2020hfs,Gautason:2020tmk,Anegawa:2020ezn,Hashimoto:2020cas,Hartman:2020swn,Hollowood:2020cou,Geng:2020qvw,Hernandez:2020nem,Li:2020ceg,Chu:2021gdb,Hu:2022ymx,He:2021mst,Ling:2020laa,Ling:2021vxe,Chen:2020tes,Balasubramanian:2020xqf,Hartman:2020khs,Almheiri:2019psy,Bousso:2020kmy,Ageev:2021ipd,Rolph:2021nan}.}:
\be\label{is} \boldsymbol{S}\left(\boldsymbol{R}\right) = \mathop {\min\;{\rm Ext}}\limits_I \left[ {{S_{\rm QFT}}\left( {A \cup I} \right) + \frac{{{\rm Area}\left( {\partial I} \right)}}{{4G_N^{\left( d \right)}}}} \right],
\ee
where boldface indicates the true von Neumann entropy of $R$ in the full quantum description (i.e., in the complete theory of quantum gravity), while the quantities which are not bolded represent the entropies calculated in the semi-classical description. This convention will always be adopted in this paper. This formula instructs us that when calculating the fine-grained entropy of a subregion $R$ of the nongravitating region, one should carefully account for the contribution of degrees of freedom in a particular region of the gravitational part, a surprising region called island. Interestingly, the work of~\cite{Chen:2020uac,Chen:2020hmv} (see also~\cite{Akal:2021foz,Suzuki:2022xwv}) shows that, in fact the island formula~(\ref{is}) in the ``black hole + radiation'' setup is equivalent to the holographic entanglement entropy formula in the AdS/BCFT setup. This allows us to investigate various problems related to the concept of ``island'' in the black hole information problem in the simple AdS/BCFT setup.

It is worth emphasizing again that since we do not yet have an authentic theory of quantum gravity, we are actually describing the behavior of systems with islands in a semi-classical picture of gravity. Nevertheless, it is interesting to investigate the entanglement details between the different parts of the system more specifically in such a semi-classical picture. However, previous work usually focused on the entanglement entropy of a subregion of the system (although see~\cite{Ageev:2021ipd,Rolph:2021nan}). In this paper, our research interest is to further study the partial entanglement entropies (PEEs) of the various parts in the subsystem in the context involving island in this semi-classical picture.

The idea of PEE came from an attempt to express the entanglement entropy in a more refined way as the sum of the contributions of each local degree of freedom in the subregion~\cite{vidal2014}. We can first define the entanglement contour ${{f_A}\left( x \right)}$ as a density function of entanglement entropy $S\left( A \right)$, satisfying
\be S\left( A \right) = \int_A {{f_A}\left( x \right)dx} ,\ee
where $x$ represents the spatial coordinates of region $A$. Then the partial entanglement entropy (PEE) ${s_A}\left( {{A_i}} \right)$ of some finite size subset ${{A_i}}$ of $A$ is defined as
\be{s_A}\left( {{A_i}} \right) \equiv \int_{{A_i}} {{f_A}\left( x \right)dx} .\ee
One can see that the PEE ${s_A}\left( {{A_i}} \right)$ captures the contribution from ${{A_i}}$ to entanglement entropy $S\left( A \right)$. The concepts of the PEE and the entanglement contour have a range of applications in studying the entanglement structures in condensed matter theory~\cite{vidal2014,Kudler-Flam:2019oru,DiGiulio:2019lpb,MacCormack:2020auw}. Moreover, they have enlightening significance in the holographic framework~\cite{Wen:2018whg,Wen:2019ubu,Wen:2018mev,Lin:2021hqs}. However, so far the fundamental definition of the PEE based on the reduced density matrix has not been established. Rather, it is required to satisfy a series of reasonable conditions according to its physical meaning~\cite{vidal2014,Wen:2019iyq}, which are, however, not sufficient to uniquely determine the PEE in general. ~\cite{Wen:2018whg, Kudler-Flam:2019oru} proposed a PEE proposal, which claims that the PEE can be obtained by an additive linear combination of subset entanglement entropies.\footnote{In fact, there exist other proposals for the PEE, see e.g.~\cite{vidal2014,Coser:2017dtb,Tonni:2017jom,Kudler-Flam:2019nhr,Wen:2018mev,Wen:2018whg,Wen:2019iyq,Han:2019scu}. Although these proposals came from different physical motivations, the PEE calculated by different approaches are highly consistent with each other~\cite{Wen:2018whg,Wen:2019iyq,Wen:2018mev,Han:2019scu,Kudler-Flam:2019nhr}.} Recently,~\cite{Lin:2021hqs} further developed the preliminary discussion of the relationship between entanglement contour and bit threads~\cite{Freedman:2016zud,Cui:2018dyq,Headrick:2017ucz} in~\cite{Kudler-Flam:2019oru}, and showed that, in the holographic framework, the PEE proposal can be naturally derived using the language of bit threads. More specifically, the PEE is explicitly identified as the flux of the $component~flow$ in a locking bit thread configuration~\cite{Headrick:2020gyq,Lin:2020yzf}.

In this paper, we will show that, using the method of calculating PEE from the viewpoint of holographic bit thread developed in~\cite{Lin:2021hqs}, it is natural to study the PEE aspects in the AdS/BCFT setup. Based on the holographic triality of AdS/BCFT setup mentioned above, in a sense we are equivalently studying the PEE aspects of the double holographic model or the more general ``brane gravity + CFT'' models. Our work reveals specific details of the entanglement between different subregions of the system involving island, which may provide further insights into the black hole information paradox. Moreover, our work shows that when considering the PEE of a subregion in a holographic BCFT in the island phase, just as one needs to be careful to distinguish between the ideas of fine-grained entropy and semi-classical entropy, we also need to redefine the fine-grained PEE, which should be distinguished from the semi-classical PEE. More specifically, we consider the appropriate definition of the fine-grained PEE of a specified subregion in BCFT in three cases, in which the subregion includes the whole boundary, no boundary at all, and only part of the boundary respectively. Interestingly, similar to the island rule of entanglement entropy~(\ref{is}), we proposed the island rules of the fined-grained PEE, which also give the prescriptions in terms of the semi-classical entropies.

The structure of this paper is as follows: In section \ref{sec2}, we review the background knowledge about AdS/BCFT setup and its holographic triality. In section \ref{sec3}, we use the locking bit thread configuration to describe the specific entanglement details between different subregions in an island phase of the AdS/BCFT setup in the semi-classical picture. In section \ref{sec4}, we define the concept of fine-grained PEE in the context involving island, and propose its island rules. The conclusion and discussion are given in section \ref{sec5}. In addition, for convenience, we include the review of bit threads in the appendix.


\section{Background review }\label{sec2}


\subsection{The basics of AdS/BCFT}\label{subsec2.1}

In the AdS/CFT correspondence, a $d+1$ dimensional AdS space (AdS$_{d+1}$) is dual to a $d$ dimensional CFT~\cite{Maldacena:1997re,Gubser:1998bc,Witten:1998qj}. In particular, the $SO (2, d)$ symmetry of AdS$_{d+1}$ geometry is equivalent to the conformal symmetry of the CFT$_d$. When there exists a $d-1$ dimensional boundary for a $d$ dimensional CFT such that the appearance of the boundary breaks $ SO(2, d)$ into $ SO(2, d-1)$, this CFT is called a BCFT. In~\cite{Takayanagi:2011zk,Fujita:2011fp}, it was proposed that AdS/CFT correspondence can be generalized to AdS/BCFT correspondence: the holographic CFT on a manifold $M$ with a boundary $\partial M$ is dual to the gravity on an asymptotic AdS space $N$ with a boundary $\partial N = M \cup Q$, where $Q$ is a codimensional-1 surface in the bulk. The $Q$ brane is also called an ETW (end-of-the world) brane or equivalently a Randall-Sundrum (RS) brane~\cite{Randall:1999ee,Randall:1999vf,Karch:2000ct} in the recent context, and it can be intuitively imagined as extended from the boundary of the BCFT, i.e., $\partial M$, see figure~\ref{f0}. In the standard AdS/CFT correspondence, the Dirichlet boundary condition is usually adopted at the AdS boundary $M$, where the CFT lives. The point of the AdS/BCFT correspondence is that, the Neumann boundary condition is imposed on another boundary $Q$ of the bulk manifold $N$, which allows the metric of $Q$ to fluctuate, and thus makes it to be dynamical. This makes it possible, as we will review below, for $Q$ brane to be described by a gravitational theory. The gravitational action of this setup is given by
\be\label{act} {I_G} = \frac{1}{{16\pi {G_N}}}\int_N {\sqrt { - g} \left( {R - 2\Lambda } \right)}  + \frac{1}{{8\pi {G_N}}}\int_Q {\sqrt { - h} \left( {K - T} \right)} ,\ee
where ${h^{ab}}$ is the induced metric and constant $T$ is the tension of the $Q$ brane, respectively, which corresponds to adding some boundary matter field whose stress-energy tensor is ${T_{ab}} =  - T{h_{ab}}$. The extrinsic curvature and its trace on $Q$ are
\be{K_{ab}} = {\nabla _a}{n_b},\quad K = {h^{ab}}{K_{ab}},\ee
where ${n^a}$ is a unit normal vector to $Q$. The Neumann boundary condition on $Q$ is computed as:
\be\label{bdy e}{K_{ab}} - K{h_{ab}} =  - T{h_{ab}}\quad  \to \quad K = \frac{d}{{d - 1}}T,\ee
which can also be called the ``boundary Einstein equation''. This equation will also determine the position of the Q brane. To see this, note that in order to maintain the $SO(2,d-1)$ symmetry of BCFT, the bulk spacetime $N$ should be foliated by AdS$_d$ slices. In fact, the bulk metric can be written as:
\be\label{fol} d{s^2} = d{\rho ^2} + {\cosh ^2}\frac{\rho }{L}ds_d^2,\ee
where the metric of AdS$_d$ can be expressed in terms of Poincare coordinates as
\be ds_d^2{\rm{ = }}{L^2}\frac{{ - d{t^2} + d{\xi ^2} + d{{\vec \chi }^2}}}{{{\xi ^2}}},\ee
and $\rho  \to   \infty $ is the AdS$_{d+1}$ boundary. Now supposing $Q$ is at the position $\rho  = {\rho _*}$, then the extrinsic curvature on $Q$ can be computed as:
\be{K_{ab}} = \frac{1}{2}\frac{{\partial {g_{ab}}}}{{\partial \rho }} = \frac{1}{L}\tanh \frac{{{\rho _*}}}{L}{h_{ab}}\ee
By~(\ref{bdy e}), the position of $Q$ brane should satisfy
\be\label{pos} T = \frac{{d - 1}}{L}\tanh \frac{{{\rho _*}}}{L}.\ee

\begin{figure}[htbp]     \begin{center}
		\includegraphics[height=7cm,clip]{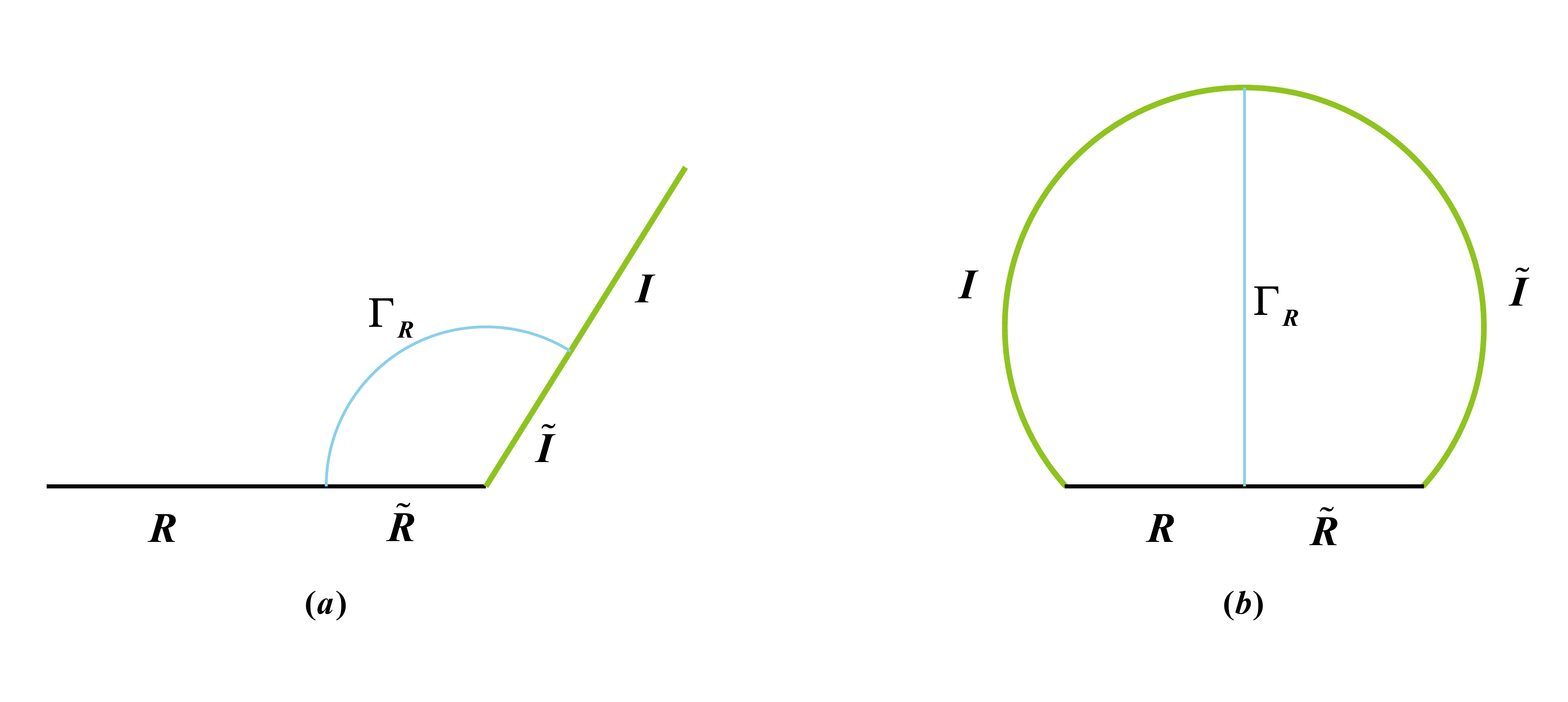}
		\caption{(a) The holographic dual of a BCFT living on the half space. (b) The holographic dual of a BCFT living on a disk. The ETW branes $Q$ are depicted in green. The RT surfaces ${\Gamma _R}$ associated with $R$ regions are depicted in blue.
		}
		\label{f1}
	\end{center}	
\end{figure}

Here we present two simple examples of the position of the $Q$ brane, which will also be used to illustrate our idea in the next section. First, using the following coordinate transformation
\be\label{tran} z = \frac{\xi }{{\cosh \frac{\rho }{L}}},\quad x = \xi \tanh \frac{\rho }{L}\ee
to make~(\ref{fol}) return to the more familiar Poincare metric
\be\label{poi} d{s^2}{\rm{ = }}{L^2}\frac{{ - d{t^2} + d{z^2} + d{x^2} + d{{\vec \chi }^2}}}{{{z^2}}}\ee
Then defining a BCFT living on the half space represented by $ x < 0$, as shown in figure~\ref{f1}(a). Then by~(\ref{pos}) and~(\ref{tran}), one can obtain that in this case the $Q$ brane is given by the plane~\footnote{The standard AdS/BCFT correspondence chooses $TL < 1$.}
\be\label{pl} z = \lambda x,\quad {\rm{where}}\;\lambda  = \sqrt {{{\left( {\frac{{d - 1}}{{LT}}} \right)}^2} - 1}.\ee

Another interesting example is the case where BCFT lives on a disk. For this, denoting the coordinates except for the radial coordinate $z$ as ${X^\mu } = \left( {\tau ,x,\vec \chi } \right)$, where $\tau  = it$ is the Euclidean time. Then applying the following conformal map (where ${c^\mu }$ are arbitrary constants)~\cite{Fujita:2011fp,Berenstein:1998ij}
\be
	{X'}_\mu  &=& \frac{{{X_\mu } + {c_\mu }{X^2}}}{{1 + 2\left( {c \cdot X} \right) + {c^2} \cdot {X^2}}}\\
	z' &=& \frac{z}{{1 + 2\left( {c \cdot X} \right) + {c^2} \cdot {X^2}}}
\ee
and performing a proper translation, then the BCFT on the half space defined by $x < 0$ can be mapped to a BCFT living on a $d$ dimensional ball with radius $ {r_B}$, defined by
\be\label{rb}{\tau ^2} + {x^2} + {\vec \chi ^2} \le r_B^2.\ee
In this way, the $Q$ brane satisfies
\be\label{qb} {\tau ^2} + {x^2} + {\vec \chi ^2} + {\left( {z - {r_B}\sinh \frac{{{\rho _*}}}{L}} \right)^2} = r_B^2{\left( {\cosh \frac{{{\rho _*}}}{L}} \right)^2},\ee
which is also a sphere, as shown in figure~\ref{f1}(b).

After determining the position of $Q$ brane,~\cite{Takayanagi:2011zk,Fujita:2011fp} proposed that the holographic entanglement entropy formula~\cite{Ryu:2006bv,Ryu:2006ef,Hubeny:2007xt} can be generalized to this AdS/BCFT setup. More specifically, considering a subsystem $R$ on a time slice of a holographic BCFT, its von Neumann entropy can be holographically computed by the following formula
\be\label{rt} \boldsymbol{S}\left(\boldsymbol{R}\right) = \mathop {\min\;{\rm Ext}}\limits_{{\Gamma _R},I} \left[ {\frac{{{\rm Area}\left( {{\Gamma _R}} \right)}}{{4G_N^{\left( {d + 1} \right)}}}} \right],\quad \partial {\Gamma _R} = \partial R \cup \partial I,\ee
where $ {\Gamma _R}$ can be called the RT surface in the AdS/BCFT setup, and $I$ is a region on the $Q$ brane, which is also named as entanglement island in the recent context~\cite{Chen:2020uac,Chen:2020hmv,Akal:2021foz,Suzuki:2022xwv}, for reasons reviewed in the next subsection. The difference between the above formula and the traditional holographic RT formula is that, now the RT surface calculating the entanglement entropy of subregion $R$ can not only be chosen as the connected extremal surface extended from the boundary of $R$ (i.e., $\partial R$), but also the disconnected type, which can anchor on the $Q$ brane. Therefore, we need to select the one with the smallest area among these two kinds of extremal surface configurations.

In figure~\ref{f1} we illustrate the two simple examples in which the island appears. From now on, for simplicity, let us focus on the two-dimensional case. In the case of half-line BCFT, we choose the subregion $R$ on a time slice as the half line defined by $x <  - l$, thus its complement is an interval $ - l \le x \le 0$, which includes the degrees of freedom of the boundary. In this case, the RT surface computing the entanglement entropy between $R$ and its complement should anchor on the ETW brane, according to which, we can holographically compute the entropy as:
\be\label{same} \boldsymbol{S}\left(\boldsymbol{R}\right)= \frac{c}{6}\log \frac{{2l}}{\varepsilon } + {S_{\rm bdy}},
\ee
where $ \varepsilon $ is the UV cut off (or lattice spacing), $c$ is the central charge of CFT, ${S_{\rm bdy}}$ is called the boundary entropy, which is holographically related with the brane tension with
\be {S_{\rm bdy}} = \frac{c}{6}{\rm arc}\tanh \left( {LT} \right).\ee

Another interesting example is that for $d=2$, the ball-shaped time slice in figure~\ref{f1}(b) becomes an interval with a length of $2{r_B}$, and contains two disconnected boundaries. If we take $R$ as half of the whole BCFT system, that is, containing only the degrees of freedom of one of the boundaries, then from symmetry we know that, in the island phase the corresponding disconnected RT surface should be exactly a straight line bisecting the whole bulk, and end on the ETW brane.

\subsection{ Holographic triality of AdS/BCFT}\label{subsec2.2}

\begin{figure}[htbp]     \begin{center}
		\includegraphics[height=11cm,clip]{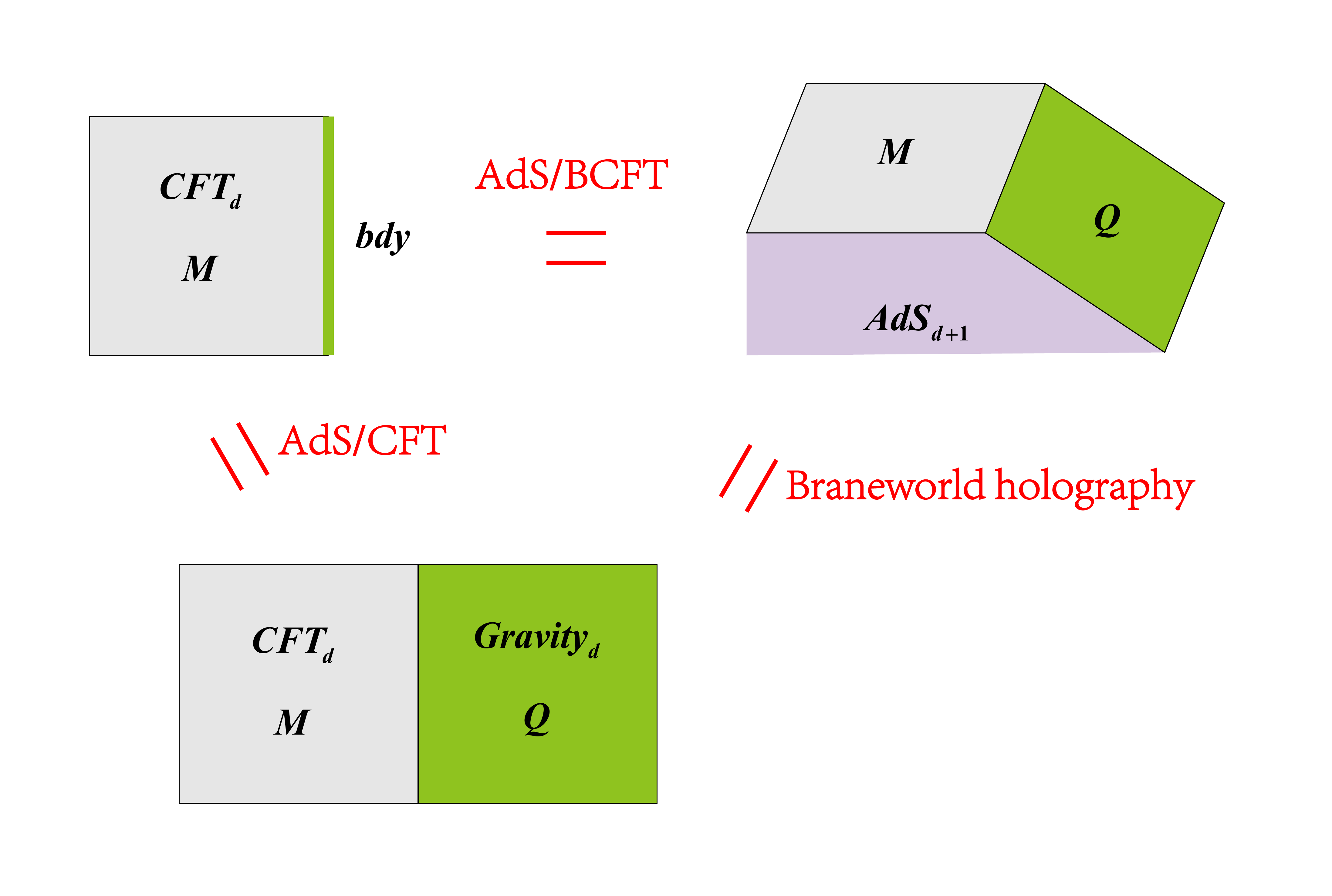}
		\caption{Holographic triality of AdS/BCFT setup. Upper left: boundary perspective, with the holographic CFT$_d$ (gray) coupled to a codimension-one conformal defect (green). Upper right: bulk gravity perspective, with an asymptotically AdS$_{d+1}$ (shaded mauve) which contains a co-dimension one ETW brane (shaded green). Bottom: brane perspective, with a non-gravitational CFT$_d$ (gray) glued to a gravity theory on the AdS$_d$ space (shaded green).}
		\label{f0}
	\end{center}	
\end{figure}

The interesting holographic triality property of the AdS/BCFT correspondence has attracted a lot of attention in the context of the black hole information problem recently~\cite{Chen:2020uac,Chen:2020hmv,Akal:2021foz,Suzuki:2022xwv}. As shown in the figure~\ref{f0}, the upper picture shows the usual AdS/BCFT correspondence, that is, we can describe a $d$ dimensional BCFT (which has a $d-1$ dimensional boundary) equivalently using an Einstein gravity on an asymptotically AdS$_{d+1}$ space containing an ETW brane (i.e., the $Q$ brane). On the other hand, as shown in the left picture, since the boundary condition in the AdS/BCFT setup is chosen to preserve the $ SO(2, d-1)$ symmetry, physics on this $d-1$ dimensional boundary can be described as a CFT$_{d - 1}$, which should correspond to an AdS$_d$ gravity from the usual AdS/CFT correspondence. According to this, one can obtain a third equivalent picture, that is, a non-gravitational CFT$_d$ is glued to a gravitational theory on the AdS$_d$ space. Naturally, this third scenario is reminiscent of the double holographic model proposed in the recent studies of the black hole information problem~\cite{Almheiri:2019hni}, in which the von Neumann entropy of the Hawking radiation $R$ can be computed by the island prescription~(\ref{is}). Interestingly, this connection can be more accurately clarified through the framework of braneworld holography~\cite{Chen:2020uac,Chen:2020hmv,Akal:2021foz,Suzuki:2022xwv}. More specifically, the Neumann boundary condition imposed on the ETW brane in the AdS/BCFT setup inspires one to relate the second picture (i.e., the bulk perspective) to the third picture (i.e., the brane perspective) by braneworld holography~\cite{Randall:1999ee,Randall:1999vf,Karch:2000ct}. Therefore, in fact the ETW brane in the AdS/BCFT setup can be regarded as a RS brane or a Karch-Randall (KR) brane brane~\cite{Randall:1999ee,Randall:1999vf,Karch:2000ct}, on which there is an effective description of a cutoff CFT coupled with gravity.

The holographic triality suggests that the island formula eq.(\ref{is}) in the ``black hole + radiation'' setup and the holographic entanglement entropy formula eq.(\ref{rt}) in the AdS/BCFT setup is actually equivalent~\cite{Chen:2020uac,Chen:2020hmv,Akal:2021foz,Suzuki:2022xwv}~\footnote{Actually, in~\cite{Chen:2020uac,Chen:2020hmv}, an additional correction was made to formula eq.(\ref{rt}), which resulted from adding an intrinsic gravity term (such as DGP gravity or JT gravity) to the brane action in the framework of braneworld holography. In this paper we focus on the traditional AdS/BCFT setup, in which we only have a tension term in eq.(\ref{act}), since this setup is sufficient to investigate the concept of island.}. In particular, we have written the holographic entanglement entropy formula in a similar way as the island rule. Therefore, these two formulas, eq.(\ref{is}) and eq.(\ref{rt}) are just two different ways of describing the same physics in the bulk perspective and the brane perspective, respectively. For the purposes of our work, the crucial point is that, from the perspective of braneworld holography, the $I$ region in eq.(\ref{rt}) in AdS/BCFT setup should be understood as an entanglement island which provides unexpected degrees of freedom for the fine-grained entropy of a subregion $R$ in the non-gravitational region.


\section{The bit thread description of AdS/BCFT setup }\label{sec3}

In this section, we will apply the locking bit thread configuration developed in~\cite{Headrick:2020gyq,Lin:2020yzf,Lin:2021hqs} to describe the details of the entanglement structure in the AdS/BCFT setup involving island. The analysis will lead to a better understanding of the reason why the region in $Q$ brane selected by the disconnected RT surface is called an ``island''. Furthermore, in the framework of bit thread, the jump of the RT surface in the phase transition can be understood in a more natural way~\cite{Freedman:2016zud}.

\begin{figure}[htbp]     \begin{center}
		\includegraphics[height=13cm,clip]{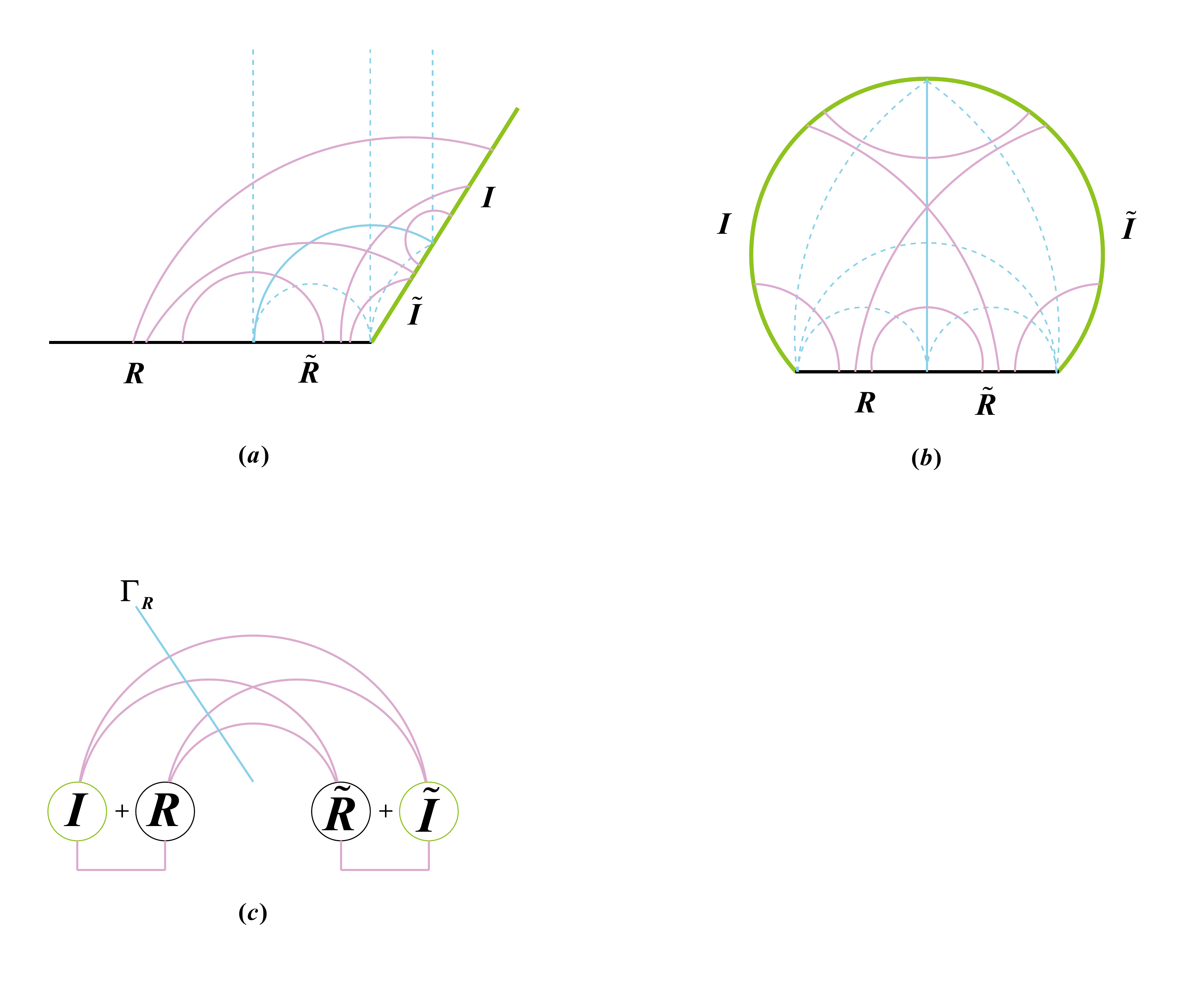}
		\caption{(a) The locking bit thread configuration characterizing the entanglement structure in the AdS/BCFT setup of figure~\ref{f1}(a). (b) The locking bit thread configuration characterizing the entanglement structure in the AdS/BCFT setup of figure~\ref{f1}(b). In both cases, there are six independent thread bundles (represented by mauve lines) and six constraints, which are the areas of a set of bulk extremal surfaces (in blue) in total. (c) Tracing the thread bundles passing through ${{\boldsymbol{\Gamma}} _{\boldsymbol{R}}}$, we find that actually there are four kinds of entanglement contributing to the von Neumann entropy $\boldsymbol{S}\left(\boldsymbol{R}\right)$, which are ${F_{R\tilde R}}$, ${F_{R\tilde I}}$, ${F_{\tilde RI}}$, and ${F_{I\tilde I}}$.}
		\label{f2}
	\end{center}	
\end{figure}

For simplicity, we will use the two simple examples shown in the previous section to illustrate our idea, as shown in figure~\ref{f2}, let us consider a fixed time slice such as $\tau  = 0$, and denote the selected subregion in the BCFT living on the holographic boundary (i.e., $z=0$) as $R$, while denote the rest of the spatial region of the BCFT system as $\tilde R$. The prescription of AdS/BCFT correspondence tells us that, when computing the entanglement entropy of $R$, one should take into account both the connected type and the disconnected type of extremal surfaces. This is because the RT formula essentially instructs us to find a minimal extremal surface in the bulk that can divide the bulk into two parts such that one part completely contacts the selected boundary subregion $R$, and the other part completely contacts the complement of $R$. Therefore, the latter type of extremal surface can also realize this idea. Let us denote the two parts of the $Q$ brane separated by this disconnected RT surface as $I$ and $\tilde I$ respectively, with $\tilde I$ adjacent to $\tilde R$.

Our motivation is that, in the framework of bit threads, when an entropy is characterized by the geometric area, it can equivalently characterized by the flux of the bit threads, namely, when an entropy can be characterized by a minimal extremal area, it implies this entropy can be characterized by the maximal flux of the bit threads, due to the max flow-min cut theorem~\cite{Freedman:2016zud,Cui:2018dyq}. In the latter formulation, however, we can trace the trajectories of the bit threads, especially their endpoints, and obtain more clear information about the entanglement structure, which is in fact related to the partial entanglement entropies. This idea is realized in the recently developed locking bit thread scheme~\cite{Headrick:2020gyq,Lin:2020yzf,Lin:2021hqs}. It has been proposed in~\cite{Lin:2021hqs} that the so-called locking bit thread configurations can be used to characterize the PEE structure between several subregions in a multipartite holographic system.

It should be pointed out that in the previous work, we were applying bit thread description in the traditional AdS/CFT framework. However, now we are facing with a holographic BCFT setup. One major difference is that in the present case, the $Q$ brane is also the boundary of the bulk spacetime, on which the Neumann boundary condition is imposed, while in the usual holographic CFT/bulk setup, the Dirichlet boundary condition is adopted at the whole boundary of the bulk. However, based on two natural reasons, the bit thread description should still be applicable to the holographic BCFT setup. In particular, the locking bit thread configuration should still be able to characterize the PEE structure of this holographic system. The first reason is due to the equivalence between the AdS/BCFT setup and the braneworld scenario. As we have reviewed, this brane perspective suggests that region $I$ is actually an entanglement island, and there exist the degrees of freedom on the $Q$ brane, which will also contribute to the entanglement entropy. In particular, it is possible to use the $Q$ brane to model the black hole spacetime~\cite{Akal:2021foz,Chen:2020hmv,Rozali:2019day}. Therefore, although the $Q$ brane is equipped with the different boundary condition, it is also a source of gravity and hence can contribute to the entanglement entropy. The second reason is that from the bulk perspective of the AdS/BCFT setup, the total contribution to the holographic entropy is only the area of a minimal surface in the classical geometric sense, unlike in the brane perspective, where one should also consider the contribution of the bulk von Neumann entropy. Therefore, the traditional bit thread formulation can always relate the area of a minimal surface with the maximal flux of the bit threads, without any new modification~\footnote{Note that this is different from recent work on discussing the island with bit threads~\cite{Agon:2021tia,Rolph:2021hgz}, where the authors needed to further modify the properties of the bit threads due to the quantum corrections to the RT formula~\cite{Faulkner:2013ana,Engelhardt:2014gca}. }.

Now following the locking bit thread scheme proposed in~\cite{Lin:2021hqs}, we can assign a locking bit thread configuration to this AdS/BCFT setup to describe the entanglement structure between the various subregions in this picture. Our setup involves four elementary regions, $R$, $ \tilde R$, $I$ and $\tilde I$. The locking thread configuration involving four elementary regions has been constructed in~\cite{Lin:2021hqs}, which involves six independent thread bundles and six constraints. Each constraint corresponds to the area of a minimal extremal surface associated with a composite region or an elementary region. As shown in figure~\ref{f2}, we use the mauve lines to represent the involved thread bundles.

Here is an important conceptual comment: From the third equivalent picture, when we refer to entropies of each specified region below, they should be carefully understood as the semi-classical entropies in the semi-classical picture (in recent literature, they correspond to the entropies not bolded in the formulas). Here the situation of the ``CFT d.o.f. + boundary d.o.f.'' composite system is similar to that of the ``radiation dof + black hole dof'' composite system. Note that when ignoring the subtle effect of the latter degrees of freedom on the von Neumann entropy (i.e., the fine-grained entropy) of a specified subregion of the former system, the von Neumann entropy of a subregion of the former system will be naively calculated by choosing its corresponding connected RT surface. Nevertheless, in the full picture, the correct semi-classical prescription to calculate this von Neumann entropy should be formula eq.(\ref{rt}). In the following we will use the notations that are not bolded to represent the semi-classical entropies (in the language of the third perspective) directly corresponding to the connected RT surfaces. For example, we denote the semi-classical entropy of $R$ simply as $ S\left( R \right) $. Instead, we will denote the true von Neumann entropy of the subregion $R$ as boldface $ \boldsymbol{S}\left(\boldsymbol{R}\right)$, which should be computed from the RT surface in prescription eq.(\ref{rt}).

We use a multiflow $ V = \left\{ {{{\vec v}_{ij}}} \right\}$ to describe this thread configuration, where each thread bundle is represented by a component flow. More concretely, each thread bundle connecting two subregions $i$ and $j$ represents the entanglement between $i$ and $j$ in physics, and is characterized by a component flow ${\vec v_{ij}}$. And the multiflow describing this locking thread configuration should satisfy the following conditions~\cite{Headrick:2020gyq,Lin:2020yzf,Lin:2021hqs}:

$(1)$. The basic conditions of multiflow:
\be\nabla  \cdot {\vec v_{ij}} &=& 0,\\
\rho (V) &\le & 1,\\
\hat{n}\cdot \vec v_{ij}|_{A_k} &=& 0,\quad({\rm for}\quad k \ne i,j) .\ee
where $\rho (V)$ is the thread density, in this case it should be defined as the number of threads per unit area intersecting a small disk, maximized over the orientation of the disk~\cite{Headrick:2020gyq}.

$(2)$. The locking conditions:

On the connected minimal extremal surface ${\gamma _i}$ associated with the specified region ${A_i}$, which is proportional to the semi-classical entropy $S\left( {{A_i}} \right)$ in our context, we have
\be{\vec v_{ij}} \bot {\gamma _i}\ee
\be{\left. {\rho (V)} \right|_{{\gamma _i}}} = 1.\ee
Next, defining
\be F{(\alpha )_{ij}} = \left| {\int_\alpha  {{{\vec v}_{ij}}} } \right| = \left| {\int_\alpha  {\sqrt h {{\hat n}_\alpha } \cdot {{\vec v}_{ij}}} } \right|\ee
to represent the value of the flux of the bit threads described by the component flow ${\vec v_{ij}}$ passing through the $\alpha $ surface, where $h$ is the determinant of the induced metric on the surface $\alpha $, and ${\hat n_\alpha }$ is the unit normal vector on surface $\alpha $. Due to the divergenceless property of bit threads, the $F{(\alpha )_{ij}}$ associated with each component flow does not depend on the surfaces the threads pass through in the bulk, and thus can be abbreviated as $ {F_{ij}} $. In the framework of the locking bit thread scheme, the flux $ {F_{ij}} $ of each thread bundle in a locking thread configuration characterizes the amount of the entanglement between the two regions it connects. Furthermore, when the entropy $Entropy\left( A \right)$ corresponding to a region $A$ can be captured by the area of an associated extremal surface $\gamma \left( A \right)$ by the ``locking rule'':
\be\label{loc} Entropy\left( A \right) = \frac{{{\rm Area}\left( {\gamma \left( A \right)} \right)}}{{4{G_N}}} = \frac{{{\rm Flux}_{\rm locking}\left( {\gamma \left( A \right)} \right)}}{{4{G_N}}},\ee
the entanglement structure of the system can be understood as follows: Tracing the starting and ending points of each thread bundle and its trajectory through the bulk, then the fluxes of those bundles just passing through the extremal surface $\gamma \left( A \right)$ is considered to make contributions to the entropy of $A$ region computed by this surface. And the values of these fluxes describe the amounts of the entanglement contributed to this entropy, see figure~\ref{f2}. Also note that we deliberately denote the entropy of $A$ as $Entropy\left( A \right)$, this is because there are two types of entropies in our analysis, one is the semi-classical entropy, denoted as $S\left( A \right)$, the other is the real von Neumann entropy of $A$ region, denoted as $\boldsymbol{S}\left(\boldsymbol{A}\right)$. And our locking rule eq.(\ref{loc}) applies to both cases.

As shown in figure~\ref{f2}, in this locking bit thread scheme, there are six independent thread bundles in total, which are ${F_{R\tilde R}}$, ${F_{RI}}$, ${F_{R\tilde I}}$, ${F_{\tilde RI}}$, ${F_{\tilde R\tilde I}}$ and ${F_{I\tilde I}}$. Our bit thread interpretation now immediately gives us some interesting conclusions. One of the most interesting questions is that, in the semi-classical picture, when an island appears in calculating the fine-grained entropy $\boldsymbol{S}\left(\boldsymbol{R}\right)$ of a subsystem $R$, which is analogy with the Hawking radiation in the black hole information problem, how exactly does entanglement between various subregions contribute to $\boldsymbol{S}\left(\boldsymbol{R}\right)$? The key point is that the RT surface corresponding to $\boldsymbol{S}\left(\boldsymbol{R}\right)$ should actually be the surface
\be{{\boldsymbol{\Gamma}} _{\boldsymbol{R}}} = \gamma \left( {\tilde R \cup \tilde I} \right) \equiv \gamma \left( {\tilde R\tilde I} \right)\ee
in the figure~\footnote{Note that in this paper we will often implicitly omit the union symbol, for example $\tilde R \cup \tilde I$ is abbreviated as $\tilde R\tilde I$.}.
Note that here ${{\boldsymbol{\Gamma}} _{\boldsymbol{R}}}$ is also denoted in boldface to indicate that it corresponds to the fine-grained entropy $\boldsymbol{S}\left(\boldsymbol{R}\right)$, while $\gamma \left( {\tilde R\tilde I} \right)$ means that in the semi-classical picture, this minimal surface appears to compute the semi-classical entropy $S\left( {\tilde R\tilde I} \right)$ of region $\tilde R \cup \tilde I$. Then tracing the thread bundles passing through ${{\boldsymbol{\Gamma}} _{\boldsymbol{R}}}$, we find that actually there are four kinds of entanglement contributing to the von Neumann entropy $\boldsymbol{S}\left(\boldsymbol{R}\right)$, which are ${F_{R\tilde R}}$, ${F_{R\tilde I}}$, ${F_{\tilde RI}}$, and ${F_{I\tilde I}}$, as shown in figure~\ref{f2}. This phenomenon actually leads to such an understanding: when calculating $\boldsymbol{S}\left(\boldsymbol{R}\right)$, it seems that we are dividing the whole system into two groups , $\left\{ {\tilde R,\tilde I} \right\}$ and $\left\{ {R,I} \right\}$, and $\boldsymbol{S}\left(\boldsymbol{R}\right)$ is essentially computing the (semi-classical) entropy between these two groups. Since ${F_{\tilde R\tilde I}}$ and ${F_{RI}}$ represent the internal entanglement in these two groups respectively, they do not contribute to the entanglement entropy between these two groups, as shown in figure~\ref{f2}(c). This interpretation is consistent with the holographic entanglement entropy formula eq.(\ref{rt}), in which $I$ and $R$ is implicitly regarded as an union. This interpretation also makes it clearer to name the $I$ region as an entanglement island on the brane in the AdS/BCFT setup. Just as in the context of the black hole information problem, as indicated by the island rule formula eq.(\ref{is}), in calculating the true von Neumann entropy of $R$ in the semi-classical picture, a somewhat unexpected island region should be regarded as an alliance with $R$ to make a contribution.

Following~\cite{Lin:2021hqs}, we can calculate the component flow fluxes (CFFs) of each thread bundle by specifying a set of constraints for this system, which correspond to the areas of the extremal surfaces. The standard choice is to take a set of extremal surfaces as shown in figure~\ref{f2}, i.e.,
\be\left\{ {\gamma \left( R \right),\;\gamma \left( {\tilde R} \right),\;\gamma \left( I \right),\;\gamma \left( {\tilde I} \right),\;\gamma \left( {\tilde R\tilde I} \right),\;\gamma \left( {R\tilde R} \right)} \right\}.\ee
In particular, $\gamma \left( {\tilde R\tilde I} \right) \equiv {{\boldsymbol{\Gamma}} _{\boldsymbol{R}}}$. Also note that a region and its complement in the entire system share the same extremal surface, for example, $\gamma \left( {\tilde R\tilde I} \right) = \gamma \left( {RI} \right)$. This is also equivalent to saying that we are constructing a locking thread configuration that can lock a specified set of boundary subregions $\left\{ {R,\;\tilde R,\;I,\;\tilde I,\;\tilde R\tilde I,\;R\tilde R} \right\}$, and thus can characterize the entanglement structure between these subregions~\cite{Lin:2020yzf,Lin:2021hqs}. Here are two conceptual comments: one may worry about whether the extremal surfaces associated with the subregions on the brane, such as $\gamma \left( {\tilde I} \right)$ in the figure has appropriate physical meaning in the semi-classical picture. Actually, in the framework of braneworld holography, this kind of extremal surfaces can indeed be associated with the entropies of the gravitational system on the brane~\cite{Emparan:2006ni,Myers:2013lva,Bianchi:2012ev,Iwashita:2006zj}. In addition, in figure~\ref{f1}(a), when the subregion is taken as the infinite half-line system, the corresponding extremal surface associated with its semi-classical entropy is an infinite straight line in the bulk.

Hence these specified extremal surfaces are associated with the semi-classical entropies $S\left( A \right)$ through the following formula:
\be\label{sa} S\left( A \right) = \frac{{{\rm Area}\left( {\gamma \left( A \right)} \right)}}{{4G_N^{\left( {d + 1} \right)}}} = \frac{{{\rm Flux}_{\rm locking}\left( {\gamma \left( A \right)} \right)}}{{4G_N^{\left( {d + 1} \right)}}}.\ee
In particular, as we have analyzed,
\be \boldsymbol{S}\left(\boldsymbol{R}\right)= \frac{{{\rm Area}\left( {{\Gamma _R}} \right)}}{{4G_N^{\left( {d + 1} \right)}}} = \frac{{{\rm Area}\left( {\gamma \left( {R \cup I} \right)} \right)}}{{4G_N^{\left( {d + 1} \right)}}} \equiv S\left( {RI} \right)\ee
Then, according to the locking conditions, we can obtain the equations depicting the entanglement structure of the island phase as follows:
\be\begin{array}{l}
	{F_{R\tilde R}} + {F_{R\tilde I}} + {F_{\tilde RI}} + {F_{I\tilde I}} = S\left( {\tilde R\tilde I} \right) = \boldsymbol{S}\left(\boldsymbol{R}\right)\\
	{F_{R\tilde R}} + {F_{RI}} + {F_{R\tilde I}} = S\left( R \right)\\
	{F_{R\tilde R}} + {F_{\tilde RI}} + {F_{\tilde R\tilde I}} = S\left( {\tilde R} \right)\\
	{F_{RI}} + {F_{\tilde RI}} + {F_{I\tilde I}} = S\left( I \right)\\
	{F_{R\tilde I}} + {F_{\tilde R\tilde I}} + {F_{I\tilde I}} = S\left( {\tilde I} \right)\\
	{F_{RI}} + {F_{R\tilde I}} + {F_{\tilde RI}} + {F_{\tilde R\tilde I}} = S\left( {R\tilde R} \right)
\end{array}\ee
To analyze the structure of the solution, let us write them in the form of matrix equation as
\be\left( {\begin{array}{*{20}{c}}
		1&1&1&0&0&0\\
		1&0&1&1&1&0\\
		1&0&0&1&0&1\\
		0&1&0&1&1&0\\
		0&0&1&0&1&1\\
		0&1&1&1&0&1
\end{array}} \right)\left[ {\begin{array}{*{20}{c}}
		{{F_{I\tilde I}}}\\
		{{F_{\tilde R\tilde I}}}\\
		{{F_{R\tilde I}}}\\
		{{F_{\tilde RI}}}\\
		{{F_{R\tilde R}}}\\
		{{F_{RI}}}
\end{array}} \right] = \left[ {\begin{array}{*{20}{c}}
		{S\left( {\tilde I} \right)}\\
		{\boldsymbol{S}\left(\boldsymbol{R}\right)}\\
		{S\left( I \right)}\\
		{S\left( {\tilde R} \right)}\\
		{S\left( R \right)}\\
		{S\left( {R\tilde R} \right)}
\end{array}} \right]\ee
The determinant of the matrix is not zero and the matrix has full rank, therefore, the solution of the equations exists and is unique. We immediately obtain the solution as
\be\label{flux}\left[ {\begin{array}{*{20}{c}}
		{{F_{I\tilde I}}}\\
		{{F_{\tilde R\tilde I}}}\\
		{{F_{R\tilde I}}}\\
		{{F_{\tilde RI}}}\\
		{{F_{R\tilde R}}}\\
		{{F_{RI}}}
\end{array}} \right] = \left( {\begin{array}{*{20}{c}}
		{\frac{1}{2}}&0&{\frac{1}{2}}&0&0&{ - \frac{1}{2}}\\
		{\frac{1}{2}}&{ - \frac{1}{2}}&0&{\frac{1}{2}}&0&0\\
		0&{\frac{1}{2}}&{ - \frac{1}{2}}&{ - \frac{1}{2}}&0&{\frac{1}{2}}\\
		{ - \frac{1}{2}}&{\frac{1}{2}}&0&0&{ - \frac{1}{2}}&{\frac{1}{2}}\\
		0&0&0&{\frac{1}{2}}&{\frac{1}{2}}&{ - \frac{1}{2}}\\
		0&{ - \frac{1}{2}}&{\frac{1}{2}}&0&{\frac{1}{2}}&0
\end{array}} \right)\left[ {\begin{array}{*{20}{c}}
		{S\left( {\tilde I} \right)}\\
		{\boldsymbol{S}\left(\boldsymbol{R}\right)}\\
		{S\left( I \right)}\\
		{S\left( {\tilde R} \right)}\\
		{S\left( R \right)}\\
		{S\left( {R\tilde R} \right)}
\end{array}} \right]\ee
or
\be\left[ {\begin{array}{*{20}{c}}
		{{F_{I\tilde I}}}\\
		{{F_{\tilde R\tilde I}}}\\
		{{F_{R\tilde I}}}\\
		{{F_{\tilde RI}}}\\
		{{F_{R\tilde R}}}\\
		{{F_{RI}}}
\end{array}} \right] = \left[ {\begin{array}{*{20}{c}}
		{{\textstyle{1 \over 2}}\left( {S\left( {\tilde I} \right) + S\left( I \right) - S\left( {R\tilde R} \right)} \right)}\\
		{{\textstyle{1 \over 2}}\left( {S\left( {\tilde I} \right) + S\left( {\tilde R} \right) - {\boldsymbol{S}\left(\boldsymbol{R}\right)}} \right)}\\
		{{\textstyle{1 \over 2}}\left( {{\boldsymbol{S}\left(\boldsymbol{R}\right)} + S\left( {R\tilde R} \right) - S\left( {\tilde R} \right) - S\left( I \right)} \right)}\\
		{{\textstyle{1 \over 2}}\left( {{\boldsymbol{S}\left(\boldsymbol{R}\right)} + S\left( {R\tilde R} \right) - S\left( {\tilde I} \right) - S\left( R \right)} \right)}\\
		{{\textstyle{1 \over 2}}\left( {S\left( {\tilde R} \right) + S\left( R \right) - S\left( {R\tilde R} \right)} \right)}\\
		{{\textstyle{1 \over 2}}\left( {S\left( R \right) + S\left( I \right) - {\boldsymbol{S}\left(\boldsymbol{R}\right)}} \right)}
\end{array}} \right]\ee
Interestingly, one can see that as expected, the correlation between $R$ region and $ \tilde R$ region is characterized by
\be{F_{R\tilde R}} = {\textstyle{1 \over 2}}\left( {S\left( {\tilde R} \right) + S\left( R \right) - S\left( {R\tilde R} \right)} \right),\ee
which is the expression of their mutual information in the semi-classical picture. Furthermore, the correlation between $R$ and $I$ in this picture can be explicitly characterized by
\be{F_{RI}} = {\textstyle{1 \over 2}}\left( {S\left( R \right) + S\left( I \right) - {\boldsymbol{S}\left(\boldsymbol{R}\right)}} \right),\ee
which is the sum of the semi-classical entropies of $R$ and $I$ minus the true von Neumann entropy of $R$. Note that this value is obviously always positive, this is because in the island phase, we have
\be {\rm Area}\left( {\gamma \left( R \right)} \right) > {\rm Area}\left( {\gamma \left( {\tilde R\tilde I} \right)} \right),\ee
and thus by~(\ref{sa}), we have $S\left( R \right) > {\boldsymbol{S}\left(\boldsymbol{R}\right)}$, and thus
\be{F_{RI}} > 0.\ee

\section{The PEE aspects of island phase}\label{sec4}

In this section, based on the ``PEE=CFF'' prescription proposed in~\cite{Lin:2021hqs}, we discuss how to appropriately redefine the concept of PEE in the context of island phase. Partial entanglement entropy (PEE) ${s_A}\left( {{A_i}} \right)$, as its name implies, measures the contribution from a part of region ${A_i}$ in $A$ to the entanglement entropy of $A$. In particular, if we take a set of ${A_i}$ satisfying they do not overlap with each other and exactly compose $A$, we expect $\sum\limits_i {{s_A}\left( {{A_i}} \right)}  = S\left( A \right)$. However, in the current context involving island, we need to be very careful with the concept of PEE, because there are two kinds of entropies. It turns out that, similarly, we should distinguish two types of partial entanglement entropy, i.e., the semi-classical PEE and the fine-grained PEE.

Although the idea of PEE is very natural, in fact the fundamental definition of the PEE based on the reduced density matrix has not been established. Rather, in general it is required to satisfy a series of reasonable conditions according to its physical meaning~\cite{vidal2014}. However, a reasonable definition of PEE can be schematically expressed as~\cite{Wen:2019iyq}
\be{s_A}\left( {{A_i}} \right) = P\left( {{A_i} \leftrightarrow \bar A} \right)\ee
This is well understood, because the contribution of ${A_i}$ to the entanglement entropy of $A$ can be naturally understood as the amount of the entanglement between ${A_i}$ and the complement of $A$, i.e., $\bar A$. In fact, in~\cite{Lin:2021hqs}, we further argue that in a locking thread configuration describing the entanglement details of the system, the PEE $P\left( {{A_i} \leftrightarrow \bar A} \right)$ can be identified with the CFF (component flow flux) ${F_{{A_i}\bar A}}$. To emphasize its physical meaning, let us denote ${F_{{A_i}\bar A}} \equiv F\left( {{A_i} \leftrightarrow \bar A} \right)$. Then the ``PEE=CFF'' prescription says
\be{s_A}\left( {{A_i}} \right) = F\left( {{A_i} \leftrightarrow \bar A} \right)\ee
This is essentially because $F\left( {{A_i} \leftrightarrow \bar A} \right)$ can indeed describe the amount of the entanglement between ${A_i}$ and the complement of $A$. Furthermore, this prescription is nicely consistent with the so-called PEE proposal, which is proposed to compute the PEE by an additive linear combination of subset entanglement entropies ~\cite{Wen:2018whg,Kudler-Flam:2019oru}.

However, we will reveal some of the quirks and subtleties of the concept of PEE in the context with island. In this section we will present several examples to illustrate the PEE aspects of the system in island phase. Interestingly, similar to the island rule of entanglement entropy in the semi-classical picture, we also obtain the island rules of PEE.

\subsection{The island rule of PEE for subregion containing the entire boundary}\label{subsec4.1}

\begin{figure}[htbp]     \begin{center}
		\includegraphics[height=7cm,clip]{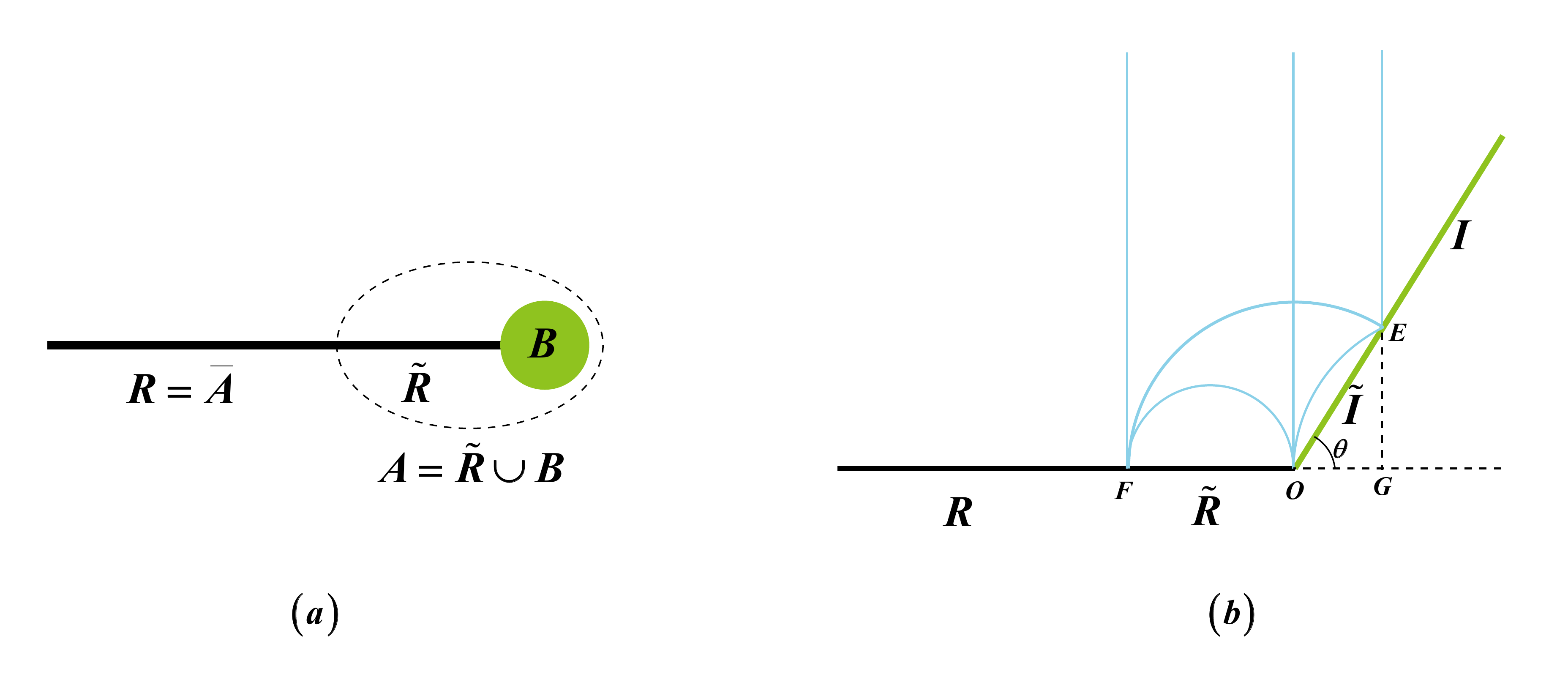}
		\caption{(a) The boundary perspective of the BCFT setup in figure~\ref{f1}(a), and we choose  $A = \tilde R \cup B$. (b) The set of the extremal surfaces (in blue) involved in the locking scheme corresponding to this setup.}
		\label{f4}
	\end{center}	
\end{figure}

\begin{figure}[htbp]     \begin{center}
		\includegraphics[height=5.5cm,clip]{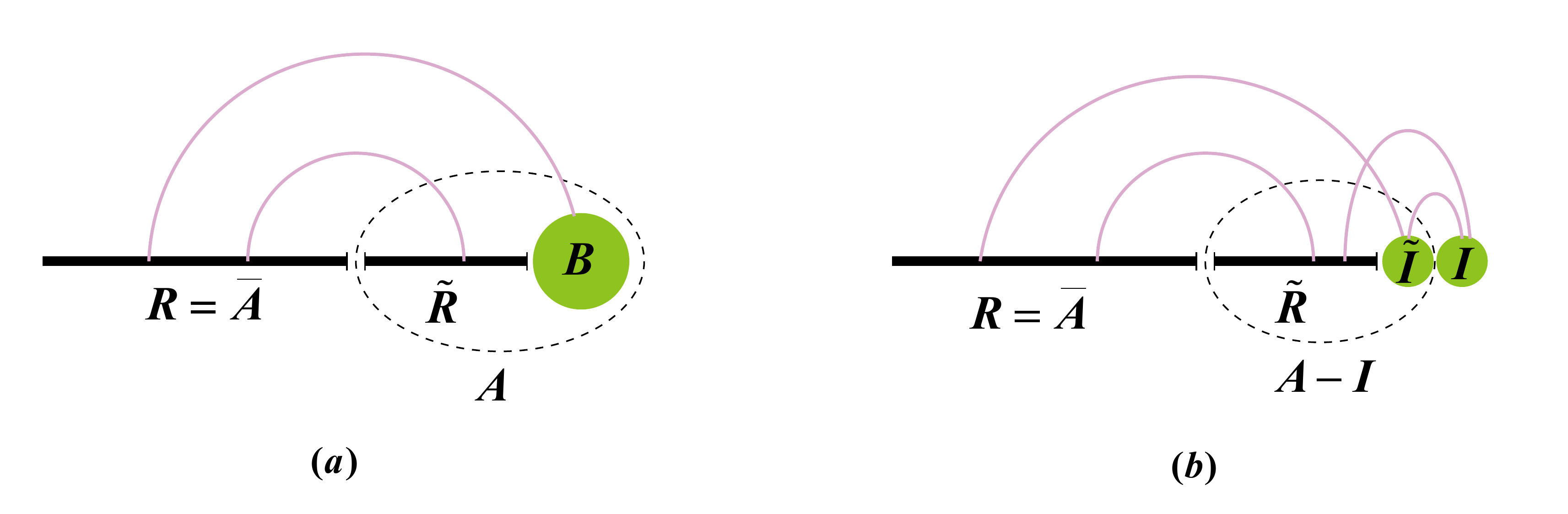}
		\caption{(a) The naive form of fine-grained PEE for subregion containing the entire boundary. (b) The island rule of fine-grained PEE for subregion containing the entire boundary.}
		\label{f5}
	\end{center}	
\end{figure}

In this subsection we consider a subregion which includes the entire boundary degrees of freedom in the holographic BCFT system. More explicitly, let us consider the BCFT setup in figure~\ref{f1}(a), but investigate it in the boundary perspective of the three equivalent scenarios, see figure~\ref{f4}(a). We will select $A$ as a subsystem containing the whole boundary degrees of freedom, which is denoted as $B$. As shown in figure~\ref{f4}, we choose $A = \tilde R \cup B$ (such that the complement of $A$ is $\bar A = R$), then we can ask how much the two components $\tilde R$ and $B$ of $A$ contribute to the von Neumann entropy of $A$, i.e., $\boldsymbol{S}\left(\boldsymbol{A}\right) =\boldsymbol{S}\left(\boldsymbol{{\bar A}}\right) =\boldsymbol{S}\left(\boldsymbol{R}\right)$ (where we have used that the whole system is in a pure state). Let us denote these two contributions as $\boldsymbol{{s_A}\left( {\tilde R} \right)}$ and $\boldsymbol{{s_A}\left( B \right)}$, and marked in bold, because they should be understood as fine-grained PEEs, which satisfy
\be\label{sum} \boldsymbol{{s_A}\left( {\tilde R} \right)} + \boldsymbol{{s_A}\left( B \right)} = \boldsymbol{S\left( A \right)}\ee

Now, naively, one may think that $\boldsymbol{{s_A}\left( {\tilde R} \right)}$ can still be expressed as the form of the entanglement between $\tilde R$ and the complement of $A$, i.e., $\bar A$, and similarly, $\boldsymbol{{s_A}\left( B \right)}$ can be expressed as the correlation between $B$ and $\bar A$, as shown in figure~\ref{f5}(a). That is, naively, one may hope the following forms are possible:
\be\label{nai}\begin{array}{l}
	\boldsymbol{{s_A}\left( {\tilde R} \right)} = P\left( {\tilde R \leftrightarrow \bar A} \right) = P\left( {\tilde R \leftrightarrow R} \right)\\
	\boldsymbol{{s_A}\left( B \right)}= P\left( {B \leftrightarrow \bar A} \right) = P\left( {B \leftrightarrow R} \right)
\end{array}\ee
However, in the analysis of the previous section, we have found that when the system is in island phase, the true RT surface calculating the true entanglement entropy between $A$ and $\bar A = R$ is actually $\boldsymbol{{\Gamma _R}}$, and if we check the thread bundles passing through $\boldsymbol{{\Gamma _R}}$, we will find that in this semi-classical picture, the contribution to $\boldsymbol{S\left( A \right)}$ collected by $\tilde R$ region is actually $F\left( {\tilde R \leftrightarrow R} \right)$ plus $F\left( {\tilde R \leftrightarrow I} \right)$, on the other hand, the contribution of $B$ to $\boldsymbol{S\left( A \right)}$ should be $F\left( {\tilde I \leftrightarrow R} \right)$ plus $F\left( {\tilde I \leftrightarrow I} \right)$, that is
\be\label{nic}\begin{array}{l}
	\boldsymbol{{s_A}\left( {\tilde R} \right)} = F\left( {\tilde R \leftrightarrow R} \right) + F\left( {\tilde R \leftrightarrow I} \right) \equiv F\left( {\tilde R \leftrightarrow R \cup I} \right)\\
	\boldsymbol{{s_A}\left( B \right) }= F\left( {\tilde I \leftrightarrow R} \right) + F\left( {\tilde I \leftrightarrow I} \right) \equiv F\left( {\tilde I \leftrightarrow R \cup I} \right)
\end{array}\ee
One can see that it does not match the naive form in eq.(\ref{nai}) in general. However, the present form in eq.(\ref{nic}) can nicely conform to eq.(\ref{sum}).

Based on the above considerations, here we propose the island rule of the fine-grained PEE in terms of semi-classical entropy, in analogy with the island rule of fine-grained entropy. More specifically, in a subsystem $A$ that includes the boundary degrees of freedom (or the degrees of freedom of brane gravity in the brane perspective), the fine-grained PEE of a spatial subregion ${{A_i}}$ that excludes the boundary degrees of freedom can be expressed as
\be\label{rule1}\boldsymbol{{s_A}\left( {{A_i}} \right)} = {s_{A - I}}\left( {{A_i}} \right),\ee
where the entropy without bold is called semi-classical PEE, which can be equivalently expressed in the traditional form, i.e.,
\be\label{semi}{s_A}\left( {{A_i}} \right) = P\left( {{A_i} \leftrightarrow \bar A} \right) = F\left( {{A_i} \leftrightarrow \bar A} \right),\ee
and the subscript $ A - I$ indicates that in the effective description, one should exclude the island part from $A$ system, such that the complement will become $\bar A \cup I$. In other words, we have
\be {s_{A - I}}\left( {{A_i}} \right) \equiv P\left( {{A_i} \leftrightarrow \bar A \cup I} \right) = F\left( {{A_i} \leftrightarrow \bar A \cup I} \right).\ee
It is easy to verify that back to our previous example, taking the subsystem as $A = \tilde R \cup B$ and $ {A_i}$ as the subregion $\tilde R$ that includes the entire spatial region except the boundary, then according to this island rule eq.(\ref{rule1}), we can recover eq.(\ref{nic}). However, we would like to point out that the island rule eq.(\ref{rule1}) here is more general, because one can take ${A_i}$ as only part of $\tilde R$. This is inspired by the entanglement structure revealed by the bit thread configuration.

On the other hand, the fine-grained PEE of the boundary degrees of freedom should be
\be\boldsymbol{{s_A}\left( B \right)} = {s_{A - I}}\left( {B - I} \right),\ee
where
\be{s_{A - I}}\left( {B - I} \right) \equiv P\left( {B - I \leftrightarrow \bar A \cup I} \right) = F\left( {\tilde I \leftrightarrow \bar A \cup I} \right),\ee
and we thus return to eq.(\ref{nic}) again.

The island rule of fine-grained PEE in this case is depicted in figure~\ref{f5}(b).

It is interesting to specifically calculate the values of the fine-grained PEEs $\boldsymbol{{s_A}\left( {\tilde R} \right)}$ and $\boldsymbol{{s_A}\left( B \right)}$, and the amount of the entanglement between $R$ and the island $I$, i.e., ${F_{RI}}$ in this case. For this purpose, we can explicitly calculate the area of each extremal surface (which corresponds to the semi-classical entropy) in the locking scheme characterizing the entanglement structure in this BCFT setup. In the Poincare metric eq.(\ref{poi}), which in the case of AdS$_3$ is
\be d{s^2}{\rm{ = }}{L^2}\frac{{ - d{t^2} + d{z^2} + d{x^2}}}{{{z^2}}},\ee
the formula for the length $d$ of the geodesic between two points $\left( {{t_1},{x_1},{z_1}} \right)$ and $\left( {{t_2},{x_2},{z_2}} \right)$ is
\be\label{ge}\frac{{{{\left( {{t_1} - {t_2}} \right)}^2} + {{\left( {{x_1} - {x_2}} \right)}^2} + {{\left( {{z_1} - {z_2}} \right)}^2}}}{{2{z_1}{z_2}}} + 1 = \cosh \frac{d}{L}\ee
As shown in figure~\ref{f4}(b), we mark the four key points as
\be\begin{array}{l}
	E = \left( {t = 0,x = l\cos \theta ,z = l\sin \theta } \right)\\
	F = \left( {t = 0,x =  - l,z = \varepsilon } \right)\\
	G = \left( {t = 0,x = l\cos \theta ,z = \varepsilon } \right)\\
	O = \left( {t = 0,x = 0,z = \varepsilon } \right)
\end{array}\ee
From eqs.(\ref{pos})(\ref{pl}), we know the value of $\theta $ in the figure as
\be\tan \theta  = \frac{z}{x} = \frac{1}{{\sinh \frac{{{\rho _ * }}}{L}}}\ee
The quantity $\theta $, or equivalently $\frac{{{\rho _ * }}}{L}$, actually characterizes the degrees of freedom of the boundary of the BCFT. To see this, rewriting eqs.(\ref{pos})(\ref{pl}), one can find that in $d=2$, we have
\be\tanh \frac{{{\rho _ * }}}{L} = LT = \tanh \frac{{6{S_{\rm bdy}}}}{c},\ee
or
\be\frac{{{\rho _ * }}}{L} = \frac{{6{S_{\rm bdy}}}}{c}.\ee
Anyway, by formula eq.(\ref{ge}), for example, we can calculate the area of the extremal surface (i.e., the length of the geodesic) corresponding to the semi-classical entropy $S\left( {\tilde I} \right)$ of $\tilde I$ by
\be\cosh \frac{{{d_{OE}}}}{L} = 1 + \frac{{{l^2}{{\cos }^2}\theta  + {{\left( {l\sin \theta  - \varepsilon } \right)}^2}}}{{2\varepsilon l\sin \theta }} = \frac{l}{{2\varepsilon \sin \theta }},\ee
thus
\be{d_{OE}} = L\ln \frac{l}{{\varepsilon \sin \theta }}.\ee
Then we have
\be S\left( {\tilde I} \right) = \frac{{{d_{OA}}}}{{4G_N^{\left( {d + 1} \right)}}} = \frac{c}{6}\ln \frac{l}{{\varepsilon \sin \theta }},\ee
where the central charge of the CFT$_2$ is
\be c = \frac{{3L}}{{2G_N^{\left( {d + 1} \right)}}},\ee
with $d=2$. Similarly, we can obtain
\be S\left( {\tilde R} \right) = \frac{{{d_{OF}}}}{{4G_N^{\left( {d + 1} \right)}}} = \frac{c}{3}\ln \frac{l}{\varepsilon },\ee
which is a famous result. And
\be\label{same2} \boldsymbol{S\left( R \right)} = S\left( {\tilde R\tilde I} \right) = \frac{{{d_{EF}}}}{{4G_N^{\left( {d + 1} \right)}}} = \frac{c}{6}\ln \frac{{2l}}{\varepsilon } + \frac{c}{6}\ln \frac{{\left( {1 + \cos \theta } \right)}}{{\sin \theta }}.\ee
One can check that it is exactly the same as eq.(\ref{same}) reviewed in section~\ref{subsec2.1}, in particular, the second term is exactly equal to ${S_{\rm bdy}}$. Next we note that the extremal surfaces corresponding to $S\left( R \right)$, $S\left( {R\tilde R} \right)$ and $S\left( I \right)$ are actually infinite straight lines in the bulk. The entropy of a half-line subsystem in the 2d CFT is also a typical result, which is
\be\label{sr} S\left( R \right) = S\left( {R\tilde R} \right) = \frac{1}{{4G_N^{\left( {d + 1} \right)}}} \cdot L\log \frac{\Lambda }{\varepsilon } = \frac{c}{6}\log \frac{\Lambda }{\varepsilon },\ee
where $\Lambda $ is the IR cut off inside the bulk. To obtain the length of the geodesic corresponding to $S\left( I \right)$, we use the formula eq.(\ref{ge}) to obtain
\be{d_{EG}} = L\ln \frac{{l\sin \theta }}{\varepsilon },\ee
thus
\be\label{si} S\left( I \right) = \frac{1}{{4G_N^{\left( {d + 1} \right)}}}\left( {L\ln \frac{\Lambda }{\varepsilon } - {d_{EG}}} \right) = \frac{c}{6}\ln \frac{\Lambda }{{l\sin \theta }}.\ee

Now it is straightforward to obtain the values of $\boldsymbol{{s_A}\left( {\tilde R} \right)}$ and $\boldsymbol{{s_A}\left( B \right)}$. By eqs.(\ref{flux})(\ref{nic}), we have
\be \boldsymbol{{s_A}\left( {\tilde R} \right)} = \frac{1}{2}\left( {\boldsymbol{S\left( R \right)} + S\left( {\tilde R} \right) - S\left( {\tilde I} \right)} \right)\ee
\be\boldsymbol{{s_A}\left( B \right)} = \frac{1}{2}\left( {\boldsymbol{S\left( R \right)} + S\left( {\tilde I} \right) - S\left( {\tilde R} \right)} \right).\ee
subsequently, we obtain
\be\boldsymbol{{s_A}\left( {\tilde R} \right) }= \frac{c}{6}\ln \frac{{2l}}{\varepsilon } + \frac{c}{{12}}\ln \frac{{1 + \cos \theta }}{2}\ee
\be\label{sab}\boldsymbol{{s_A}\left( B \right)} = \frac{c}{{12}}\ln \frac{{2\left( {1 + \cos \theta } \right)}}{{{{\sin }^2}\theta }}.\ee
Moreover, we can specifically obtain the amount of the entanglement between $R$ and the island $I$ as
\be{F_{RI}} = {\textstyle{1 \over 2}}\left( {S\left( R \right) + S\left( I \right) - \boldsymbol{S\left( R \right)}} \right) = \frac{c}{{12}}\ln \frac{{{\Lambda ^2}}}{{2{l^2}\left( {1 + \cos \theta } \right)}}\ee

\subsection{The island rule of PEE for subregion containing no boundary}\label{subsec4.2}
We can also consider the case that the subsystem does not contain the boundary degrees of freedom. For example, we still consider the setup in figure~\ref{f4}, but focus on the subsystem $R$. The concept of PEE becomes even more interesting. In general, in the existing literature about PEE, when we focus on the entanglement entropy of a subsystem $R$ (which will be analogous to the specified radiation region in the context of black hole information problem), one will take $X$ as a subset ${R_i} \subset R$ of $R$ region and then talk about the contribution of $X$ to this entropy, i.e., the PEE ${s_R}\left( X \right)$. However, in the context with island, in fact $X$ can also be taken as the island $I$. This is simply because we have
\be\label{enc} \boldsymbol{S\left( R \right)} = {F_{R\tilde R}} + {F_{R\tilde I}} + {F_{\tilde RI}} + {F_{I\tilde I}}.\ee
Therefore, at least formally we have the following island rule
\be\label{rule2} \boldsymbol{{s_R}\left( I \right)} = {F_{\tilde RI}} + {F_{I\tilde I}} = F\left( {I \leftrightarrow \tilde R \cup \tilde I} \right) \equiv {s_{R + I}}\left( I \right),\ee
where the subscript $R+I$ represents the union of $R$ and $I$, and in the last equation, we have applied the formula eq.(\ref{semi}). This result is a bit of a surprise, but it reflects the spirit of the island rules. In the context with island, we know that the semi-classical description has some quirk, that is, at least superficially, there exists an unexpected region called island contributing to the true entanglement entropy of $R$. Actually, our work is to quantify the amount of this apparent contribution of island for the true entanglement entropy of $R$.

Similar to the formula eq.(\ref{rule1}), for a subregion ${R_i}$ in $R$, we also have
\be\boldsymbol{{s_R}\left( {{R_i}} \right)} = F\left( {{R_i} \leftrightarrow \tilde R \cup \tilde I} \right) \equiv {s_{R + I}}\left( {{R_i}} \right).\ee
Again, this is due to the entanglement structure inspired by the bit thread configuration. In particular, if we take ${R_i} = R$, we obtain
\be\label{rule22}\boldsymbol{{s_R}\left( R \right)} = {F_{R\tilde R}} + {F_{R\tilde I}}\ee
Indeed, in the semi-classical picture, it does not make the full contribution to the fine-grained entropy of $R$.

Similarly, from eqs.(\ref{flux})(\ref{rule2})(\ref{rule22}), we can express the fine-grained PEEs as the linear combination of the semi-classical entropies as follows:
\be\boldsymbol{{s_R}\left( I \right)} = \frac{1}{2}\left( {\boldsymbol{S\left( R \right)} + S\left( I \right) - S\left( R \right)} \right)\ee
\be\boldsymbol{{s_R}\left( R \right)} = \frac{1}{2}\left( {\boldsymbol{S\left( R \right)} + S\left( R \right) - S\left( I \right)} \right).
\ee
Then substituting results in eqs.(\ref{same2})(\ref{sr})(\ref{si}),  we obtain
\be\boldsymbol{{s_R}\left( I \right)} = \frac{c}{{12}}\ln \frac{{2\left( {1 + \cos \theta } \right)}}{{{{\sin }^2}\theta }} = \frac{1}{2}{S_{\rm bdy}} + \frac{c}{{12}}\ln \frac{2}{{\sin \theta }}.\ee
Interestingly, this is a finite value. In addition, we have
\be\boldsymbol{{s_R}\left( R \right)} = \frac{c}{{12}}\ln \frac{{2l\left( {1 + \cos \theta } \right)}}{{{\varepsilon ^2}}}.\ee

\subsection{The island rule of PEE for subregion containing part of boundary}\label{subsec4.3}

\begin{figure}[htbp]     \begin{center}
		\includegraphics[height=7cm,clip]{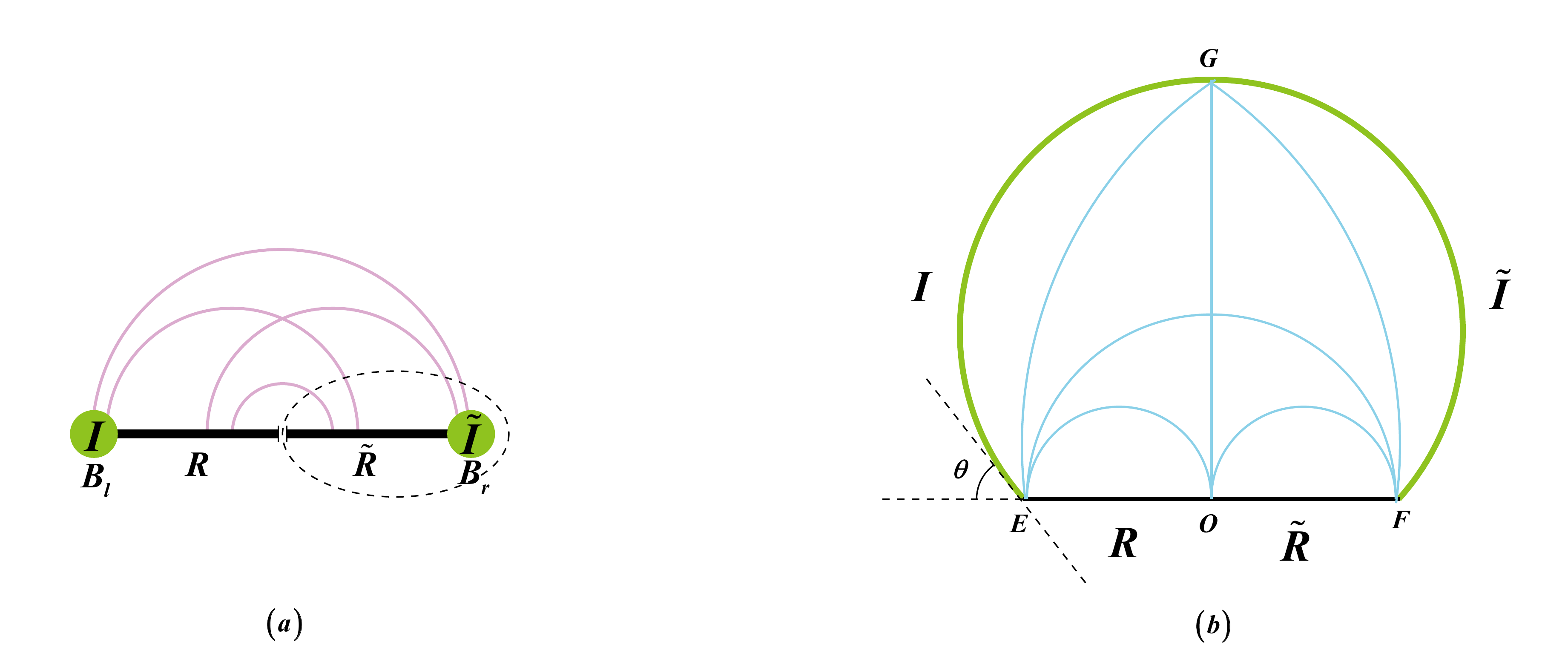}
		\caption{(a) The island rule of fine-grained PEE for subregion containing only a part of boundary degrees of freedom. The setup is the same as in figure~\ref{f1}(b). (b) The set of the extremal surfaces involved in the locking scheme corresponding to this setup.}
		\label{f6}
	\end{center}	
\end{figure}

As reviewed in subsection~\ref{subsec2.1}, we can consider another interesting case, in which the selected subregion contains only a part of boundary degrees of freedom. As shown in figure~\ref{f6}(a), let us consider the simplest symmetrical case to illustrate our idea. We now investigate the system in the boundary perspective, and denote the boundary degrees of freedom in the left and right as ${B_l}$ and ${B_r}$ respectively. We then focus on the subsystem $A = \tilde R \cup {B_r}$. One interesting thing is that in this case the boundary degrees of freedom have been split into two parts as $B = {B_l} \cup {B_r}$, and it turns out that this division of degrees of freedom is exactly corresponding to the division of the degrees of freedom on the $Q$ brane as $Q = I \cup \tilde I$. The reason is as follows.

Similar to the previous example, we have
\be\label{pa} \boldsymbol{S\left( A \right)} = {F_{R\tilde R}} + {F_{R\tilde I}} + {F_{\tilde RI}} + {F_{I\tilde I}},\ee
and we want to construct the fine-grained PEE satisfying
\be\boldsymbol{{s_A}\left( {\tilde R} \right)} + \boldsymbol{{s_A}\left( {\tilde I} \right)} = \boldsymbol{S\left( A \right)}.\ee
It is clear that from eq.(\ref{pa}), the contribution of $\tilde R$ region and ${B_r}$ to the fine-grained entropy of $A$ should be respectively
\be\label{rule3}\begin{array}{l}
	\boldsymbol{{s_A}\left( {\tilde R} \right)} = F\left( {\tilde R \leftrightarrow R} \right) + F\left( {\tilde R \leftrightarrow I} \right) \equiv F\left( {\tilde R \leftrightarrow R \cup I} \right)\\
	\boldsymbol{{s_A}\left( {{B_r}} \right)} = F\left( {\tilde I \leftrightarrow R} \right) + F\left( {\tilde I \leftrightarrow I} \right) \equiv F\left( {\tilde I \leftrightarrow R \cup I} \right)
\end{array}\ee
Then it is natural to make an identification with
\be{B_l} \sim I,\quad {B_r} \sim \tilde I,\ee
and by the semi-classical ``PEE=CFF'' prescription, we have simply
\be\boldsymbol{{s_A}\left( {\tilde R} \right)} = {s_A}\left( {\tilde R} \right)\ee
\be\boldsymbol{{s_A}\left( {{B_r}} \right) }= {s_A}\left( {{B_r}} \right).\ee
Similarly, we calculate the areas of the extremal surfaces involved in the locking scheme characterizing the entanglement structure of this setup in the semi-classical picture. As shown in figure~\ref{f6}(b), the coordinates of the key points can be obtained from the trajectories of the BCFT boundary eq.(\ref{rb}) and the brane eq.(\ref{qb}):
\be\begin{array}{l}
	E = \left( {t = 0,x =  - {r_B},z = \varepsilon } \right)\\
	F = \left( {t = 0,x = {r_B},z = \varepsilon } \right)\\
	G = \left( {t = 0,x = 0,z = {r_B}\left( {\sinh \frac{{{\rho _ * }}}{L} + \cosh \frac{{{\rho _ * }}}{L}} \right)} \right) = \left( {t = 0,x = 0,z = {r_B}\frac{{\cos \theta  + 1}}{{\sin \theta }}} \right)\\
	O = \left( {t = 0,x = 0,z = \varepsilon } \right)
\end{array},\ee
where again we use $\tan \theta  = \frac{1}{{\sinh \frac{{{\rho _ * }}}{L}}}$, and $\theta $ is again the angle between the brane and the BCFT system at the boundary. Then from eq.(\ref{ge}),
\be\cosh \frac{{{d_{EG}}}}{L} = \frac{{r_B^2 + {{\left( {{r_B}\frac{{\cos \theta  + 1}}{{\sin \theta }} - \varepsilon } \right)}^2}}}{{2\varepsilon {r_B}\frac{{\cos \theta  + 1}}{{\sin \theta }}}} + 1,\ee
thus
\be{d_{EG}} = {d_{FG}} = L\ln \frac{{2{r_B}}}{{\varepsilon \sin \theta }},\ee
thus
\be S\left( I \right) = S\left( {\tilde I} \right) = \frac{c}{6}\ln \frac{{2{r_B}}}{{\varepsilon \sin \theta }}.\ee
Similarly,
\be \boldsymbol{S\left( R \right)} = S\left( {RI} \right) = \frac{c}{6}\ln \frac{{{r_B}\left( {\cos \theta  + 1} \right)}}{{\varepsilon \sin \theta }} = \frac{c}{6}\ln \frac{{{r_B}}}{\varepsilon } + \frac{c}{6}\frac{{{\rho _ * }}}{L} \equiv \frac{c}{6}\ln \frac{{{r_B}}}{\varepsilon } + {S_{\rm bdy}},\ee
and
\be S\left( R \right) = S\left( {\tilde R} \right) = \frac{c}{3}\ln \frac{{{r_B}}}{\varepsilon },\ee
\be S\left( {R\tilde R} \right) = \frac{c}{3}\ln \frac{{2{r_B}}}{\varepsilon },\ee
which correspond to the familiar semicircular geodesics. Then with eqs.(\ref{rule3})(\ref{flux}), we obtain in this case
\be\boldsymbol{{s_A}\left( {\tilde R} \right)} = \frac{1}{2}\left( {\boldsymbol{S\left( R \right)} + S\left( {\tilde R} \right) - S\left( {\tilde I} \right)} \right),\ee
\be\boldsymbol{{s_A}\left( {{B_r}} \right)} = \frac{1}{2}\left( {\boldsymbol{S\left( R \right)} + S\left( {\tilde I} \right) - S\left( {\tilde R} \right)} \right),\ee
thus
\be\boldsymbol{{s_A}\left( {\tilde R} \right)} = \frac{c}{6}\ln \frac{{{r_B}}}{\varepsilon } + \frac{c}{{12}}\ln \frac{{1 + \cos \theta }}{2}\ee
\be\boldsymbol{{s_A}\left( {{B_r}} \right)} = \frac{c}{{12}}\ln \frac{{2\left( {1 + \cos \theta } \right)}}{{{{\sin }^2}\theta }}.\ee
Interestingly, we find that the value of $\boldsymbol{{s_A}\left( {{B_r}} \right)}$ is the same as that of the $\boldsymbol{{s_A}\left( B \right)}$ in the first case, see eq.(\ref{sab}). And
\be{F_{RI}} = {\textstyle{1 \over 2}}\left( {S\left( R \right) + S\left( I \right) - \boldsymbol{S\left( R \right)}} \right) = \frac{c}{{12}}\ln \frac{{2{r_B}^2}}{{{\varepsilon ^2}\left( {1 + \cos \theta } \right)}} = \frac{c}{6}\ln \frac{{{r_B}}}{\varepsilon } + \frac{c}{{12}}\ln \frac{2}{{1 + \cos \theta }}.\ee

\subsection{Insights into the black hole information problem}\label{subsec4.4}

\begin{figure}[htbp]     \begin{center}
		\includegraphics[height=5cm,clip]{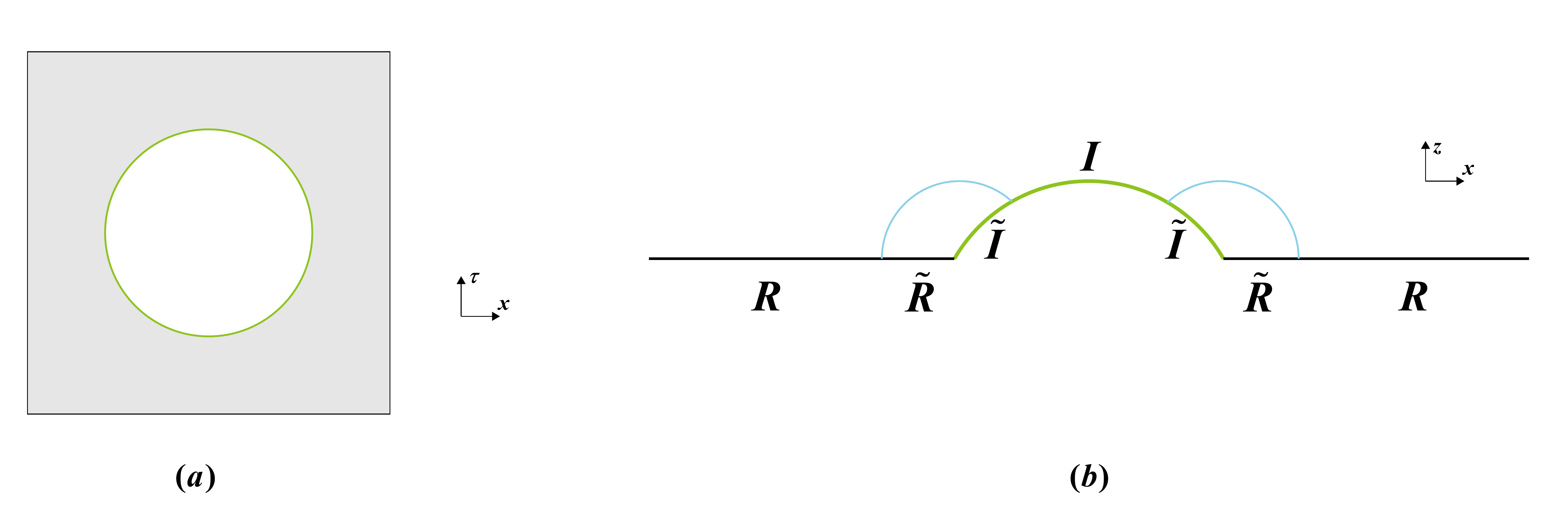}
		\caption{(a) A holographic BCFT setup (from the boundary perspective) that models a two-sided 2d black hole (in green) coupled to a pair of symmetrical auxiliary radiation systems (in grey). (b) The RT surface (in blue) calculating the true entanglement entropy of $R$ can anchor on the ETW brane (in green, which simulates a black hole) to form an island.}
		\label{f7}
	\end{center}	
\end{figure}

In this subsection we discuss the further applications of our work to the black hole information problem. As mentioned earlier, the AdS/BCFT setup can actually model the ``black hole + radiation system'' in the context of this problem. One of the most direct and interesting examples is in~\cite{Rozali:2019day} (see also~\cite{Sully:2020pza}), which simulated a two-sided black hole coupled to an auxiliary radiation system by a holographic BCFT system in the thermofield double state by applying the AdS/BCFT correspondence. This setup is similar to the one in figure~\ref{f1}(b) in which the boundary of a half plane is mapped to the boundary of a disk, except that now we map the half plane on which the BCFT lives to the Euclidean plane with the disk removed, as shown in figure~\ref{f7}(a). It was argued that in the limit that the number of local degrees of freedom on the boundary of this BCFT is large compared to the number of local degrees of freedom in this bulk CFT itself, the ETW brane extending from the boundary of the disk can simulate a black hole because the brane itself has causal horizons. In this way this setup models a two-sided 2d black hole coupled to a pair of symmetrical auxiliary radiation systems.

Note that this system is not an evaporating black hole, but one the auxiliary radiation system has the same temperature as the black hole such that the two systems are in equilibrium. Furthermore, in a particular conformal frame, this system has a static energy density. However, the calculation in~\cite{Rozali:2019day} showed that, for a subsystem $R = \left( { - \infty ,\; - {x_0}} \right] \cup \left[ {{x_0},\;\infty } \right)$ consisting of the union of  two symmetric half-lines in each CFT, the entanglement entropy still evolves with time and undergoes a typical phase transition characterized by Page curve, similar to the ones discussed in~\cite{Almheiri:2019hni,Almheiri:2019psf,Penington:2019npb}. This phase transition is essentially because in AdS/BCFT correspondence, the RT surface calculating the true entanglement entropy of $R$ can be in an island phase, i.e., the RT surface can anchor on the ETW brane, as shown in figure~\ref{f7}(b).

Our work provides a clearer picture for the island phase of entanglement entropy in such ``black hole + radiation'' systems. In figure~\ref{f7}(b), although there is no net energy exchange between the black hole and the radiation system, the information from black hole ``escapes'' (or is ``encoded'') into the radiation system $R$. As can be seen from eq.(\ref{enc}), actually the island region $I$ inside the black hole also contributes to the fine-grained entropy of subsystem $R$. In other words, it is not only $\boldsymbol{{s_R}\left( {{R_i}} \right)}$, but also $\boldsymbol{{s_R}\left( I \right)}$ contribute to $\boldsymbol{S\left( R \right)}$. Another power of bit thread interpretation is that it provides a continuous viewpoint of the phase transition between the two types of RT surface configurations involving in the calculation of the von Neumann entropy of subsystem $R$. Actually, one of the initial motivations of bit thread formulation is that it is possible to describe this kind of apparent jump of the RT surface configurations in terms of bit threads~\cite{Freedman:2016zud}. For example, considering a combined system $AB$ of two separated regions $A$ and $B$, the RT surface calculating its entanglement entropy can jump under continuous deformations of $AB$. However, unlike the minimal surfaces, the threads do not jump under these continuous deformations. That is, in the framework of bit threads, no matter what kind of RT surface configuration is presented in the holographic bulk, the locking bit thread configuration describing the entanglement of the system actually has the same structure. Therefore, the change of the thread configuration is continuous at the critical point of ``phase transition''.

An interesting problem is to compute the evolution of the fine-grained PEE of this ``black hole + radiation'' setup with time by using our prescription in terms of the combination of various semi-classical entropies. However, this may involve the covariant form of RT formula~\cite{Hubeny:2007xt,Dong:2016hjy}, and the covariant form of bit thread formulation is still under development~\cite{Freedman:2016zud}, so a more careful investigation of this issue would be left for further work. On the other hand, it is also possible to argue that our prescription is independent of the bit thread formulation and is still correct even for time-dependent situations.

\section{Conclusions and discussions}\label{sec5}
In this paper, we investigate the PEE aspects of the holographic BCFT setups in the context with entanglement islands, by combining two interesting dualities developed recently. The first duality is the triality of the AdS/BCFT setup inspired by the recent research of the black hole information paradox~\cite{Chen:2020uac,Chen:2020hmv,Akal:2021foz,Suzuki:2022xwv}, in which a $d$ dimensional BCFT can not only be described using an Einstein gravity on an asymptotically AdS$_{d+1}$ space containing an ETW brane by the usual AdS/BCFT correspondence, but also can be viewed from the so-called brane perspective through braneworld holography, that is, described as a non-gravitational CFT$_d$ glued to a gravity theory on the AdS$_d$ space. In particular, it is possible to design the holographic BCFT setup such that the effective theory on the brane is describing the black hole physics. Another duality is the ``PEE=CFF'' prescription proposed in~\cite{Lin:2021hqs}, where in the framework of  holographic bit threads, the partial entanglement entropy (PEE) is explicitly identified as the component flow flux (CFF) in a locking bit thread configuration. Combining these two insights, we study the entanglement details between a set of specified subsystems in the presence of the entanglement island.

Our work is mutually beneficial to both sides. On the one hand, inspired by the recent study on the black hole information problem, we study the PEE aspects in the holographic BCFT setups. In the previous literature, PEE is often determined by the so-called PEE proposal. However, our study shows that, when considering the PEE of a subsystem in a holographic BCFT, just as we need to distinguish between fine-grained entropy and semi-classical entropy carefully, we should also distinguish the fine-grained PEE from the semi-classical PEE. Moreover, the definition of fine-grained PEE varies subtly depending on whether the subsystem contains all or only part of the boundary degrees of freedom. We propose the island rules for the fine-grained PEE, which instruct us to calculate the fine-grained PEE in terms of the combination of the semi-classical entropies. On the other hand, our study provides a bit thread description for the AdS/BCFT setup, which characterizes the entanglement details between the different subregions in a system with the island, and this in turn provides further insights into the context of the black hole information problem. Our description presents the detailed entanglement between the different parts of the gravity-bath system in the semi-classical picture, which helps us to further understand the concept of ``island''. This can be seen most clearly in the fine-grained PEE $\boldsymbol{{s_R}\left( I \right)}$, which characterizes the amount of the contribution of the island region to the fine-grained entropy $\boldsymbol{S\left( R \right)}$ of a subregion $R$ in the bath. Furthermore, the picture of bit threads makes it possible to view the phase transition of the entanglement entropy during the black hole evaporation in a continuous way, because instead of using the jumping RT surface to characterize this transition, the change in the locking bit thread configuration is continuous.

For the future, it is interesting to consider more concrete AdS/BCFT setups modelling the ``black hole + radiation'' systems~\cite{Chen:2020hmv,Akal:2021foz,Rozali:2019day} and calculate the PEEs of the specified subregions therein, then it is possible to investigate the phase transition of the PEE itself. Furthermore, it is also interesting to consider the PEE aspects of the case where the intrinsic gravity such DGP gravity is added to the ETW brane in the traditional AdS/BCFT setup, which will lead to a brane correction to the holographic RT formula~\cite{Chen:2020uac}.

\section*{Acknowledgement}
This project was supported by the National Natural Science Foundation of China (No.~11675272) and (No.~12105113).

\begin{appendix}

\section{Introduction to bit threads and locking thread configurations}\label{app1}

Bit threads are unoriented bulk curves that end on the boundary and subject to the rule that the thread density is less than 1 everywhere~\cite{Freedman:2016zud,Cui:2018dyq,Headrick:2017ucz}. In particular, this thread density bound implies that the number of threads passing through the minimal surface $\gamma \left( A \right)$ that separates a boundary subregion $A$ and its complement ${{A_{\rm{c}}}}$ cannot exceed its area Area$\left( {\gamma \left( A \right)} \right)$, hence the flux of bit threads Flux$\left( A \right)$ connecting $A$ and its complement ${{A_{\rm{c}}}}$ does not exceed Area$\left( {\gamma \left( A \right)} \right)$:
\be\label{bound} {\rm Flux}\left( A \right) \le {\rm Area}\left( {\gamma \left( A \right)} \right) .\ee
Borrowing terminology from the theory of flows on networks, a thread configuration is said to $lock$ the region $A$ when the bound~(\ref{bound}) is saturated. Actually, this bound is tight: for any $A$, there does exist a locking thread configuration satisfying:
\be {\rm Flux}_{{\rm{locking}}}\left( A \right) = {\rm Area}\left( {\gamma \left( A \right)} \right) .\ee
This theorem is known as max flow-min cut theorem (see \cite{Headrick:2017ucz} and references therein), that is, the maximal flux of bit threads (over all possible bit thread configurations) through a boundary subregion $A$ is equal to the area of the bulk minimal surface $\gamma \left( A \right)$ homologous to $A$. Therefore, the famous RT formula which relates the entanglement entropy of a boundary subregion $A$ and the area of the bulk minimal extremal surface ${\gamma \left( A \right)}$ homologous to A:
\be S\left( A \right) = \frac{{\rm Area}\left(\gamma (A)\right)}{4G_N}\ee
can be expressed in another way, that is, the entropy of a boundary subregion $A$ is proportional to the flux of the locking thread configuration passing through $A$:
\be S\left( A \right) = \frac{{\rm Flux}_{\rm locking}\left( A \right)}{4G_N}.\ee

When the bit threads are required to be locally parallel, one can use the language of $flow$ to describe the behavior of bit threads conveniently in mathematics, that is, using a vector field $\vec v$ to describe the bit threads, just as using the magnetic field $\vec B$ to describe the magnetic field lines. The difference is that for the latter we regard the magnetic field itself as the more fundamental concept, while for the former we consider the threads to be more fundamental. The constraints on the bit threads can then be expressed as the requirements for the flow $\vec v$ as follows,
\be\nabla  \cdot \vec v &=& 0,\\
\rho \left( {\vec v} \right) &\equiv & \left| {\vec v} \right| \le 1.\ee

For situations involving more than one pair of boundary subregions, the concept of $thread~ bundles$ is also useful. The threads in each thread bundle are required to connect only a specified pair of boundary subregions, while still satisfy the constraints of bit threads. Specifically, one can use a set of vector fields ${\vec v_{ij}}$ to represent each thread bundle connecting the ${A_i}$ region and ${A_j}$ region respectively. The set $V$ of vector fields ${\vec v_{ij}}$ is referred to as a $multiflow$, and each ${\vec v_{ij}}$ is called a $component~ flow$, satisfying (Note that in the present paper we will define ${\vec v_{ij}}$ only with $i < j$ for convenience, which is slightly different from (but equivalent) convention adopted in~\cite{Cui:2018dyq}, where the fields ${\vec v_{ij}}$ were also defined for $i \ge j$, but with the constraint ${{\vec v}_{ji}} =  - {{\vec v}_{ij}}$)
\be\nabla  \cdot {\vec v_{ij}} &=& 0,\\
\rho (V) &\le & 1,\\
\hat{n}\cdot \vec v_{ij}|_{A_k} &=& 0,\quad({\rm for}\quad k \ne i,j) .\ee
It is worth noting that, since in the situation of multiflows, the threads are not necessarily locally parallel, there are various natural ways the density can be defined, and therefore bounded. It turns out that different definitions of the thread density will actually affect the ability of a thread configuration to lock a set of boundary regions.

Consider a $d$-dimensional compact Riemannian manifold-with-boundary $M$, for example, it can be a time slice of AdS$_{d + 1}$ spacetime, and then divide its boundary system $\partial M$ into adjacent non-overlapping subregions ${A_1}, \ldots ,{A_n}$, which are referred to as $elementary~regions$, satisfying ${A_i} \cap {A_j} = \emptyset $, $\mathop  \cup \limits_{i = 1}^n {A_i} = \partial M$. Accordingly, a $composite~ region$ is defined as the union of some certain elementary regions.

For a single boundary subregion $A$, the max flow-min cut theorem directly indicates that one can find a thread configuration that can lock the specified boundary subregion (and its complement simultaneously). In other words, there exist thread configurations that can lock the set of two $elementary$ $regions$ $I = \left\{ {{A_1},{A_{\rm{c}}}} \right\}$, and there is typically an infinite number of choices. However, one can further ask, can we find a locking thread configuration that can lock an arbitrary specified set of subregions simultaneously? The question becomes very nontrivial. Broadly speaking, it depends not only on the relative spatial position relations between these specified subregions, but also on the properties we assign to the bit threads, in particular, the precise definition of the thread density bound. Recently, the authors in~\cite{Headrick:2020gyq} investigated this issue in great detail. They proposed and proved several theorems on the existence of locking thread configurations in various situations in terms of the language of elementary regions and composite regions defined above. Details can be found in the original literature~\cite{Headrick:2020gyq}.

The holographic bit thread formulation has helped uncover aspects of holographic entanglement and the related quantities, for the recent developments of bit threads see e.g.~\cite{Lin:2021hqs,Lin:2020yzf,Headrick:2020gyq,Agon:2021tia,Rolph:2021hgz,Chen:2018ywy,Hubeny:2018bri,Agon:2018lwq,Du:2019emy,Bao:2019wcf,Harper:2019lff,Agon:2019qgh,Du:2019vwh,Agon:2020mvu,Bao:2020uku,Pedraza:2021fgp,Pedraza:2021mkh,Harper:2018sdd}. In particular, in~\cite{Lin:2020yzf}, by matching the locking thread configurations with the so-called OSED (one-shot entanglement distillation) tensor network developed in~\cite{Bao:2018pvs,Bao:2019fpq,Lin2020,Yu:2020qup}, the locking thread configurations are argued to provide an interesting picture of reconstructing the spacetime, i.e., the emergence of spacetime can be regarded as the reorganization of the boundary degree of freedom through the entanglement distillation.

\end{appendix}



\end{document}